\documentclass[twocolumn,showpacs,preprintnumbers,amsmath,amssymb]{revtex4}

\usepackage{graphicx}
\usepackage{dcolumn}
\usepackage{bm}
\usepackage{epsfig}

\begin{document}

\preprint{JLAB-THY-05-385}

\title{High-enegy effective action from scattering of QCD shock waves }

\author{Ian Balitsky}
\affiliation{
Physics Dept., ODU, Norfolk VA 23529, \\
and \\
Theory Group, Jlab, 12000 Jeffeson Ave, Newport News, VA 23606
}
\email{balitsky@jlab.org}

\date{\today}

\begin{abstract}
At high energies, the relevant degrees of freedom are Wilson lines - infinite gauge links 
ordered along straight lines collinear to the velocities of colliding particles.
The effective action for these Wilson lines is determined by the scattering of QCD shock waves. 
I develop the symmetric expansion of the effective action in powers of strength of one of the shock waves and
calculate the leading term of the series.  
The corresponding first-order effective action, symmetric with respect to projectile and target, includes 
both up and down fan diagrams and pomeron loops.

\end{abstract}


\maketitle

\section{\label{sec:in}Introduction }

It is widely believed that the relevant degrees of freedom for the description of 
high-energy scattering in QCD are Wilson lines - infinite straight-line gauge factors. 
An argument in favor of this goes as follows \cite{mobzor}.
As a reslut of a high-energy collision, we have a shower of produced particles in the whole range of rapidity between the target and the spectator. Let us demonstrate that
the interaction of gluons  with a different rapidity is described in terms of Wilson lines. 
Consider the fast particle interacting with some slow gluons.  This particle moves along its classical  trajectory - a straight line collinear to the velocity, and the only effect
of the slow gluons is the phase factor $P\exp\{ig\int dx_\mu A^\mu\}$ ordered along the straight-line classical path (here $A_\mu$ describes the slow gluons). This picture is reciprocal -
in the rest frame of fast particles the fast and slow gluons trade places: 
 former slow gluons move very fast 
so their propagator reduces to a Wilson line made from the (former) fast gluons.
We see that 
the particles with different rapidities perceive each other as Wilson lines and
therefore these lines must be the relevant degrees of freedom for high-energy scattering. The goal
of this approach is to rewrite the original functional integral over gluons (and quarks)
as a $2+1$ theory with the effective action written in terms  of the dynamical Wilson lines.

\begin{figure}
\includegraphics[width=0.47\textwidth]{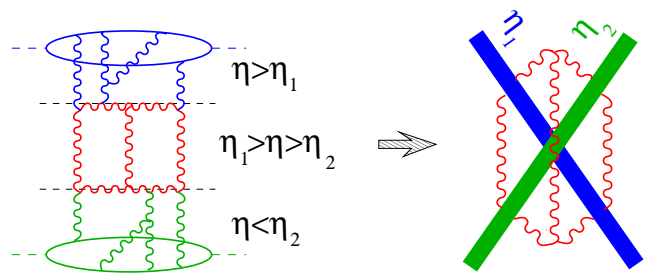}
\caption{High-energy effective action as an amplitude of the collision of two shock waves\label{fig:figura1}.}
\end{figure}

For a given interval of rapidity, the effective action is an amplitude of scattering of two QCD shock waves, see Fig. \ref{fig:figura1}. Indeed, let us integrate over the gluons in this interval of rapidity $\eta_1>\eta>\eta_2$ leaving the gluons with $\eta>\eta_1$ (the ``right-movers') and with $\eta<\eta_2$ (the ``left-movers') intact  (to be integrated over later).
Due to the Lorentz contraction, the left-moving and the right-moving gluons shrink to
the two gluon ``pancakes'' or shock waves. The result of the integration over 
the rapidities $\eta_1>\eta>\eta_2$ is the effective action which depends on the
Wilson lines made from the left-and right-movers. 

Due the parton saturation at high energies\cite{GLR, muchu,mu90}, the characteristic scale 
of the transverse momenta in hadron-hadron collisions is 
$Q_s\sim e^{c\eta}$ \cite{mu99,mabraun,iancu02,tolpa} and therefore
the collision of QCD shock waves can be treated using semiclassical methods \cite{mvmodel}. 
Within the semiclassical approach,
 the problem of scatering of two shock waves can be reduced to the solution of classical
 YM equations with sources being the shock waves \cite{nncoll} (see also \cite{prd99}). 
At present, these equations have not been solved. 
There are two approaches discussed in current
literature: numerical simulations \cite{krasvenu} and expansion in the strength of 
one of the shock waves. The collision of a weak and a strong shock waves corresponds to the deep inelastic scattering from a
nucleus (and scattering of two strong shock waves describes a nucleus-nucleus collision). The first term of the expansion of the strength of one of the waves  was calculated in a number of papers \cite{kovmu99,kop,wkov}. 
Recently, the classical field was
calculated up to the second order in a weak source \cite{prd04}. 
I will use some formulas of Ref. \cite{prd04}, although the main result for the effective action 
will be derived independently. The obtained effective action, symmetric with respect to 
projectile and target, has a ``built-in'' projectile-target duality (which is a highly nontrivial
property of the light-cone Hamiltonian in the framework of the Hamiltonian approach\cite{smith,larecent,kl05,
mu05,murecent}). 
In terms of Feynman diagrams the effective action includes  both ``up'' and ``down'' fan diagrams 
and therefore it describes pomeron loops which are a topic of intensive discussion 
in the current literature\cite{levin,smith,larecent,kl05,
mu05,murecent}.

The paper is organized as follows.
Sec. 2 is devoted to the rapidity factorization which is the starting point of the shock-wave approach.  In Sect. 3 I define the high-energy effective action
as a scattering amplitude of QCD shock waves and develop the expansion in commutators of Wilson lines.
In Sec. 4  I find the effective action for a given (infinitesimal) range of rapidity in the leading 
order in this expansion.
The corresponding functional integral over the dynamical Wilson-line variables is constructed in Sec. 5.  
 The explicit form of the first-order classical fields 
created by the collision of two shock waves  is presented in the Appendix.


\section{Rapidity factorization}

The main technical tool of the shock-wave approach to the high-energy scattering is
 the rapidity factorization
developed in \cite{prl,prd99}.
Consider a functional integral for the typical scattering amplitude 
\begin{equation}
\int\! DA J(p_A)J(p_B)
J(-p'_A)J(-p'_B)~e^{iS(A)}
\label{typi}
\end{equation}
where the currents $J(p_A)$ and $J(p_B)$ describe the two colliding 
particles (say, photons). 

Throughout the paper, we use Sudakov variables
\begin{equation}
k~=~\alpha p_1+\beta p_2+k_\perp
\label{sudakov}
\end{equation}
and the notations 
\begin{eqnarray}
&&\hspace{0mm}
x_\bullet=p_1^\mu x_\mu=\sqrt{s\over 2}x^-,~~~~x^-={1\over\sqrt{2}}(x^0-x^3)
\nonumber\\
&&\hspace{0mm}
x_\ast=p_2^\mu x_\mu=\sqrt{s\over 2}x^+,~~~~x^+={1\over\sqrt{2}}(x^0+x^3)
\label{astbullet}
\end{eqnarray}
Here $p_1$ and $p_2$ are the light-like vectors close to $p_A$ and $p_B$:
$p_A=p_1+{p_A^2\over s}p_2$, $p_B=p_2+{p_B^2\over s}p_1$. 

\begin{figure}
\includegraphics[width=0.27\textwidth]{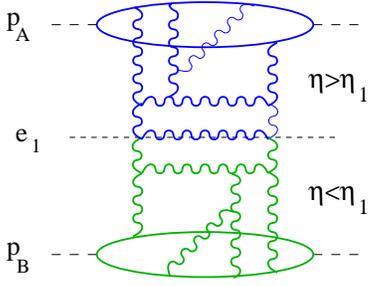}
\caption{Rapidity factorization\label{fig:figura2}.}
\end{figure}

\begin{figure}
\includegraphics[width=0.47\textwidth]{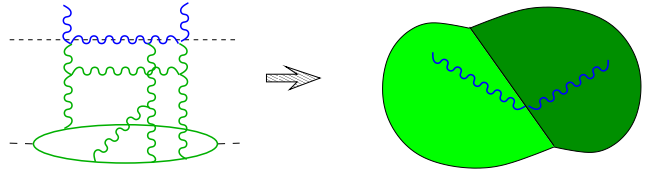}
\caption{Shock wave in the temporal gauge\label{fig:figura6}.}
\end{figure}

Let us take some ``rapidity
divide'' $\eta_1$ such that $\eta_A>\eta_1>\eta_B$ and
integrate first over the gluons with the rapidity $\eta>\eta_1$, see Fig. \ref{fig:figura2}.
 From the viewpont 
of such particles, the fields with $\eta<\eta_1$ shrink to a shock wave so 
the result of the integration is presented by Feynman diagrams in the
shock-wave background. With the LLA accuracy, in the Feynman integrals over the
gluons with $\eta>\eta_1$ one can set $\eta_1\rightarrow -\infty$ (replace the
``rapidity divide'' vector $e_1=p_1+e^{-\eta_1}p_2$ by the light-like vector $p_2$) so 
the shock wave is infinitely thin and light-like. In the covariant gauge, 
this shock-wave has the
only non-vanishing component $A_\bullet$ which is concentrated near $x_\ast=0$.
In order to write down factorization we need to rewrite the shock wave in 
the temporal gauge $A_0=0$. In such gauge the most general form of a shock-wave background is (see Fig. \ref{fig:figura6})
\begin{eqnarray}
&&\hspace{0mm}
A^i={\cal U}_1^i\theta(x_\ast)+{\cal U}_2^i\theta(-x_\ast),~~~A_\bullet=A_\ast=0
\label{shok}
\end{eqnarray}
where 
\begin{equation}
{\cal U}_1^i~=~U_1^\dagger{i\over g}\partial_iU_1,~~~~~
{\cal U}_2^i~=~U_1^\dagger{i\over g}\partial_iU_2
\label{defui}
\end{equation}
are the pure gauge fields (filling the
half-spaces $x_\ast<0$ and $x_\ast<0$ ).
There is a redundant gauge symmetry 
\begin{equation}
U_1(x_\perp)\rightarrow U_1(x_\perp)\Omega(x_\perp),~~~~
U_2(x_\perp)\rightarrow U_2(x_\perp)\Omega(x_\perp)
\label{redunu}
\end{equation}
related to the fact that gauge invariant objects like the color dipole 
\begin{eqnarray}
&&\hspace{-4mm}
{\rm Tr}\{[\infty p_2,-\infty p_2]_x[x_\perp-\infty p_2, y_\perp-\infty p_2]
[-\infty p_2,\infty p_2]_y
\nonumber\\
&&\hspace{-4mm}
\times~
[y_\perp+\infty p_2, x_\perp+\infty p_2]\}~\simeq~ {\rm Tr}\{
U_{1x}U^\dagger_{2x}U_{2y}U^\dagger_{1y}\}
\label{dipole}
\end{eqnarray}
depend only on the product $U_{1z}U^\dagger_{2z}$. In papers \cite{prd99,mobzor}
this symmetry was used to gauge away $U_2$ and simplify the shock wave to 
$A_i={\cal U}_i\theta(x_\ast)$
while in Ref. \cite{prd04} the opposite case $U_1=0$ ($A_i={\cal U}_i\theta(-x_\ast)$) was considered. 
In the present paper we keep this gauge freedom - as we shall see below it simplifies the effective action for the Wilson-line integral.

The generating functional for the Green functions in the Eq. (\ref{shok}) 
has the form (cf.  \cite{mobzor}) 
\begin{eqnarray}
&&\hspace{-13mm}
\int\! \!DA J(p_A)J(-p'_A)~e^{iS(A)
+i\!\int\! d^2z_\perp 
 (0,F_{\ast i},0)^a_z({\cal U}_1^{ai}-{\cal U}_2^{ai})_z}
 \label{generfun}
\end{eqnarray}
where ($F_{ei}\equiv e^\mu F_{\mu i}$ etc.)
\begin{eqnarray}
&&\hspace{-3mm}
(0,F_{e i},0)_z\equiv
\!\int_{-\infty}^\infty\! \!\!\!du~
[0,ue]_zF_{e i}(ue+z_\perp)[ue,0]_z
~\nonumber\\
&&\hspace{-3mm} 
=~[0,\infty e]_z(i{\partial\over\partial z^i}+gA_i(\infty e+z_\perp))[\infty e,0]_z
\label{deff}\\
&&\hspace{-3mm} 
-~[0,-\infty e]_z(i{\partial\over\partial z^i}+gA_i(-\infty e+z_\perp))[-\infty e,0]_z
\nonumber
\end{eqnarray}
 and $(0,F_{\mu i},0)^a\equiv 2{\rm tr}~t^a(0,F_{\mu i},0)$. (Throughout the paper, the sum over the Latin indices $i,j...$ runs over the two 
transverse components while the sum over Greek indices runs over the 
four components as usual).

It is easy to see that the functional integral (\ref{generfun})
generates Green functions in the Eq. (\ref{shok}) 
background. Indeed, let us choose the gauge $A_\ast=0$ for
simplicity. In this gauge,  
$(0,F_{\ast i},0)^a=A_i(\infty p_2+z_\perp)-A_i(-\infty p_2+z_\perp)$
so the functional integral (\ref{generfun})  takes the form 
\begin{eqnarray}
&&\hspace{0mm}
\int\!DA~J(p_A)J(-p'_A)
\\
&&\hspace{0mm}
\times~e^{iS(A)
+i\!\int\! d^2z_\perp 
 (A_i(\infty p_2+z_\perp)-A_i(-\infty p_2+z_\perp))^a({\cal U}_1^{ai} -{\cal U}_2^{ai})_z}  
\nonumber
\end{eqnarray}
Let us now shift the fields $A_i\rightarrow A_i+\bar{A}_i$ and  where
 $\bar{A}^i={\cal U}_1^i\theta(x_\ast)+{\cal U}_2^i\theta(-x_\ast)$. 
The only non-zero components of the classical field strength in our case are 
$F_{\bullet i}=({\cal U}_{1i}-{\cal U}_{2i})\delta({2\over s}x_\ast)$ 
so we get 
\begin{eqnarray}
&&\hspace{-5mm}S(A+\bar{A})~=~{2\over s}\!\int d^4zD^i\bar{F}_{i\bullet}A^\ast
-{2\over s}\int\! d^2z_\perp dz_\bullet
\nonumber\\
&&\hspace{-5mm}\times~\left.
A^i\bar{F}_{\bullet i}\right|^{x_\ast=\infty}_{x_\ast=-\infty}
+{1\over 2} A^\mu(\bar{D}^2g_{\mu\nu}-2i\bar{F}_{\mu\nu})A^\nu +...
\label{afteshift}
\end{eqnarray}
 In the $A_\ast=0$ gauge the first term in the r.h.s. of
Eq. (\ref{afteshift}) vanishes while the second term cancels with the corresponding
contribution $\sim - (A_i(\infty p_2+z_\perp)-A_i(-\infty p_2+z_\perp))^a{\cal U}^{ai}$ 
coming from the source in Eq. (\ref{generfun}). We obtain
\begin{eqnarray}
&&\hspace{0mm}
\int\!DA~J(p_A)J(-p'_A)
\\
&&\hspace{0mm}
\times~e^{iS(A)
+i\!\int\! d^2z_\perp 
 (A_i(\infty p_2+z_\perp)-A_i(-\infty p_2+z_\perp))^a({\cal U}_1^{ai} -{\cal U}_2^{ai})_z}  
  \nonumber\\
&&\hspace{0mm}
=~\int\!DA~J(p_A)J(-p'_A)e^{{i\over 2}\!\int\! d^2zA^\mu(\bar{D}^2g_\mu\nu
-2i\bar{F}_{\mu\nu})A^\nu}
\nonumber
\end{eqnarray}
which gives the Green functions in the 
Eq. (\ref{shok}) background.

To complete the factorization formula one needs to integrate 
over the remaining fields
with rapidities $\eta<\eta_1$: 
\begin{eqnarray}
&&\hspace{-3mm}
\int\!D{\cal A} DA {\cal J}(p_A){\cal J}(p_B)e^{-iS({\cal A})}~e^{iS(A)}
J(-p_A)J(-p_B)
\nonumber\\
&&\hspace{-3mm} 
=~\int\! DA J(p_A)J(-p'_A)\int\! DB J(p_B)J(-p'_B)
\nonumber\\
&&\hspace{-3mm}
\times~e^{iS(A)+iS(B)
+i\!\int\! d^2z_\perp 
 (0,F_{e_1 i},0)^a_z
 (0,G_{e_1 i},0)^a_z }
 \label{faktor}
\end{eqnarray}
where  the Wilson-line operators $(0,F_{e_{_1} i},0)^a_z$ and $(0,G_{e_{_1} i},0)^a_z$
are the operators (\ref{deff}) made from $A$ and $B$ fields, respectively.
As discussed in \cite{prl,prd99,mobzor,baba03}, the slope of Wilson lines 
is determined by the 
``rapidity divide'' vector $e_{\eta_1}=p_1+e^{-\eta_1}p_2$. (From the wiewpoint
of $A$ fields, the slope $e_1$ can be replaced by $p_2$ with power accuracy so we
recover the generating functional (\ref{generfun}) with 
$(0,G_{\ast i},0)={\cal U}_{1i}-{\cal U}_{2i}$).

\section{Scattering of OCD shock waves}
\subsection{Efffective action as a shock-wave scattering amplitude}

\begin{figure}
\includegraphics[width=0.47\textwidth]{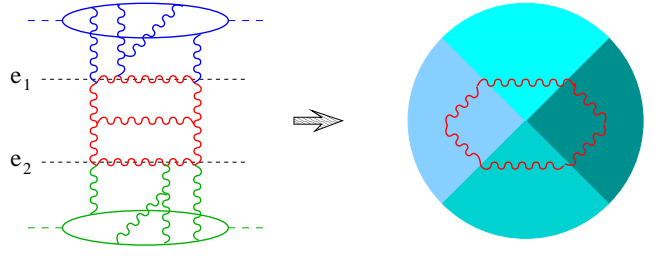}
\caption{Effective action as a scattering of two shock waves\label{fig:figura5}.}
\end{figure}

In this section we define the scattering of the shock waves 
using the rapidity factorization
developed above.
Applying the factorization formula (\ref{faktor}) two times, one gets (see Fig. \ref{fig:figura5}):
\begin{eqnarray}
&&\hspace{-3mm}
\int\!DA~J(p_A)J(p_B)J(-p'_A)J(-p'_B)~e^{iS(A)}~=
\label{fak2times}\\
\nonumber\\
&&\hspace{-3mm} \int\! DAJ(p_A)J(-p'_A)
e^{iS(A)}
\!\int\!DB~J(p_B)J(-p'_B)~e^{iS(B)}
\nonumber\\
&&\hspace{-3mm} 
\times\!\int\! DC
 \exp\Big[iS(C)+i\!\int\! d^2z_\perp \Big\{
[0,A_{e_{_1} i},0]^a_z
 [0,C_{e_{_1} i},0]^a_z
\nonumber\\
&&\hspace{27mm} 
+~(0,C_{e_{_2} i},0)^a_z
 (0,B_{e_{_2} i},0)^a_z\Big\}\Big]
 \nonumber
\end{eqnarray}
where the slope is $e_1=p_1+e^{-\eta_1}p_2$ for the $[...]$  Wilson lines
and $e_2=p_1+e^{-\eta_2}p_2$ for the $(...)$ ones.

The functional integral over the central range of rapidity 
$\eta_1>\eta>\eta_2$ is determined by the integral over $C$ field with the sources
\begin{eqnarray}
&&\hspace{-4mm}
(0,A_{e_{_1} i},0)_z
=[0,\infty e_{_1}]_z(i\partial_i+gA_i(\infty e_1+z_\perp))[\infty e_{_1},0]_z
\nonumber\\
&&\hspace{-4mm} 
-~[0,-\infty e_{_1}]_z(i\partial_i+gA_i(-\infty e_1+z_\perp))[-\infty e_{_1},0]_z
\nonumber\\
&&\hspace{-4mm}
(0,B_{e_{_2} i},0)_z
=[0,\infty e_{_2}]_z(i\partial_i+gB_i(\infty e_2+z_\perp))[\infty e_{_2},0]_z
\nonumber\\
&&\hspace{-4mm} 
-~[0,-\infty e_{_2}]_z(i\partial_i+gB_i(-\infty e_2+z_\perp))[-\infty e_{_2},0]_z
\end{eqnarray}
 made from ``external'' $A$ and $B$ fields.
Since $A_i(\pm\infty)$ is a pure gauge these sources can be represented as a difference 
of a pure-gauge fields $(0,A_{e_{_1} i},0)_z=({\cal V}_1^{ai}-{\cal V}_2^{ai})_z$ and 
$(0,B_{e_{_2} i},0)_z=({\cal U}_1^{ai}-{\cal U}_2^{ai})_z$ where 
\begin{eqnarray}
&&\hspace{-8mm}
V_{1,2}(z_\perp)=[0,\pm\infty e_1]_z[\pm\infty e_1+z_\perp,\pm\infty e_1 +\infty e_\perp]
\nonumber\\
&&\hspace{-8mm} 
U_{1,2}(z_\perp)=[0,\pm\infty e_2]_z[\pm\infty e_2+z_\perp,\pm\infty e_2 +\infty e_\perp]
\label{sorses}
\end{eqnarray}
Since there is no field strength $F_{\mu\nu}$ at infinite time the direction of $e_\perp$ does not
matter.

The result of the integration over the $C$ field is an effective action for the $\eta_1>\eta>\eta_2$ interval 
of rapidity
\begin{eqnarray}
&&\hspace{-3mm}
e^{iS_{\rm eff}(V_1,V_2,U_1,U_2;\eta_1-\eta_2)}
~ \label{effactdef}\\
&&\hspace{-3mm}=\int\!\! DC
 \exp\Big[iS(C)+
i \!\int\! d^2z_\perp \Big\{  ({\cal V}_1^{ai}-{\cal V}_2^{ai})_z
 [0,C_{e_{_1} i},0]^a_z
 \nonumber\\
&&\hspace{27mm} +~
({\cal U}_1^{ai}-{\cal U}_2^{ai})_z
 (0,C_{e_{_2} i},0)^a_z
  \Big\}\Big].
 \nonumber
\end{eqnarray}
One can interpret Eq. (\ref{effactdef}) as an effective action for scattering of two QCD shock waves
defined by the sources (\ref{sorses}). Note that the effective action 
$iS_{\rm eff}(V_1,V_2,U_1,U_2;\eta_1-\eta_2)$ defined by Eq. (\ref{effactdef}) is invariant under the redundant gauge transformations  (\ref{redunu})
\begin{eqnarray}
&&\hspace{-3mm}
U_{1(2)}(x_\perp)\rightarrow U_{1(2)}(x_\perp)\Omega(x_\perp), 
 \nonumber\\
&&\hspace{-3mm} 
V_{1(2)}(x_\perp)\rightarrow V_{1(2)}(x_\perp)\Omega'(x_\perp)
 \label{redun}
\end{eqnarray}
since this transformation can be absorbed by a gauge rotation of the $C$ fields
\begin{eqnarray}
&&\hspace{-3mm}
C_\mu\rightarrow
 \Omega^\dagger(x_\perp,\ln{x_\ast\over x_\bullet}) C_\mu  
 \Omega(x_\perp,\ln{x_\ast\over x_\bullet})
 \nonumber\\
&&\hspace{-3mm} 
+~
{i\over g}\Omega^\dagger(x_\perp,\ln{x_\ast\over x_\bullet}) \partial_\mu  \Omega(x_\perp,\ln{x_\ast\over x_\bullet})
 \label{redunrot}
\end{eqnarray}
where $ \Omega(x_\perp,\ln{x_\ast\over x_\bullet})$ is an arbitrary  $SU_3$ matrix
satisfying the conditions
$ \Omega^\dagger(x_\perp,\eta_1) =\Omega(x_\perp)$ and 
$ \Omega^\dagger(x_\perp,\eta_2) =\Omega'(x_\perp)$. 

With a power accuracy $O(m^2/s)$, we can replace $e_1$ by $p_1$ and $e_2$ by $p_2$ :
\begin{eqnarray}
&&\hspace{-5mm}
e^{iS_{\rm eff}(V_1,V_2,U_1,U_2;\eta_1-\eta_2)}~
 \label{integc}\\
&&\hspace{-3mm}
=~\int\!\! DC
 \exp\Big\{iS(C) +
i \!\int\! d^2z_\perp \Big[  ({\cal V}_1^{ai}-{\cal V}_2^{ai})_z
 [0,C_{\bullet i},0]^a_z
\nonumber\\
&&\hspace{30mm} +~
({\cal U}_1^{ai}-{\cal U}_2^{ai})_z
 (0,C_{\ast i},0)^a_z
 \Big] \Big\}
\nonumber
\end{eqnarray}

The saddle point of the functional integral (\ref{integc}) is determined by the 
classical equations
\begin{eqnarray}
&&\hspace{-0mm}
{\delta\over\delta C_\mu^a}\Big\{S(C) 
 +
i \!\int\! d^2z_\perp \Big[  ({\cal V}_1^{ai}-{\cal V}_2^{ai})_z
 [0,C_{\bullet i},0]^a_z
\nonumber\\
&&\hspace{10mm} +~
({\cal U}_1^{ai}-{\cal U}_2^{ai})_z
 (0,C_{\ast i},0)^a_z
 \Big] \Big\}~=~0
\label{cliqs}
\end{eqnarray}
At present it is not known how to solve this equations 
(for the numerical approach see \cite{krasvenu}). In the next section 
we will develop a ``perturbation theory'' in powers of the parameter 
$[U,V]\sim g^2[{\cal U}_i,{\cal V}_j]$. Note that the 
conventional perturbation theory 
corresponds to the case when ${\cal U}_i,{\cal V}_i\sim 1$ while the semiclassical QCD is relevant
when the fields are large (${\cal U}_i$ and/or ${\cal V}_i\sim{1\over g}$).

\subsection{ \label{sect:expaw}Expansion in commutators of Wilson lines}
The effective action is defined by the functional integral (\ref{integc})
(hereafter we switch
back to the usual notation $A_\mu$ for the integration variable and $F_{\mu\nu}$ for
the field strength)

\begin{eqnarray}
&&\hspace{-5mm}
e^{iS_{\rm eff}(V_1,V_2,U_1,U_2;\eta_1-\eta_2)}~
\nonumber\\
&&\hspace{-3mm}
=~\int\!\! DA
 \exp\Big(iS(A) +
i \!\int\! d^2z_\perp \Big\{ ({\cal V}_1^{ai}-{\cal V}_2^{ai})_z
 [0,F_{\bullet i},0]^a_z
\nonumber\\
&&\hspace{-3mm} +~
({\cal U}_1^{ai}-{\cal U}_2^{ai})_z
 (0,F_{\ast i},0)^a_z
  \Big\}\Big)
 \label{mastegral}
\end{eqnarray}
Taken separately, the sources $\sim{\cal U}_i$ create
a shock wave ${\cal U}_{1i}\theta(x_\ast)+{\cal U}_{2i}\theta(-x_\ast)$ and those
$\sim {\cal V}_i$ create ${\cal V}_{1i}\theta(x_\bullet)+{\cal V}_{2i}\theta(-x_\bullet)$ 
In QED, the two sources $ {\cal U}_i$ and $ {\cal V}_i$ do not interact 
(in the leading order in $\alpha$) so the sum of 
the two shock waves
\begin{eqnarray}
&&\hspace{-5mm}\bar{A}_i^{(0)}={\cal U}_{1i}\theta(x_\ast)+
{\cal U}_{2i}\theta(-x_\ast)+{\cal V}_{1i}\theta(x_\bullet)+{\cal V}_{2i}\theta(-x_\bullet)~,
\nonumber\\
&&\hspace{-3mm}
\bar{A}_\bullet^{(0)}=\bar{A}_\ast^{(0)}=0
 \label{shoksum}
\end{eqnarray}
is a classical solution to the set of equations (\ref{cliqs}). In QCD, the
interaction between these two sources is described by the commutator $g[{\cal U}_i,{\cal V}_k]$ 
(the coupling constant $g$  corresponds to the three-gluon vertex).  
The straightforward approach is to take the trial
configuration in the form of a sum of the two shock waves
and expand the ``deviation'' of the full QCD solution from
the QED-type ansatz (\ref{shoksum}) in powers of commutators $[U,V]$. 
This is done  rigorously in \cite{prd04} and the relevant formulas are presented in the Appendix. Here we will use a slightly different zero-order approximation (cf. \cite{mobzor}) which leads to 
same results in a more streamlined way at a price of some uncertainties (like $\theta(0)$) which, however, do not contribute to the effective action in the
leading order. 

 Let us consider the behavior of the solution
of the YM equations at, say, $x_0\rightarrow \infty$, $x_3$ fixed (in the forward quadrant of the space). Since there is no field strength at $t\rightarrow\infty$, the field must be a pure gauge. As demonstrated in Ref. \cite{prd04},  this pure-gauge field has the form of a sum of the shock waves plus a  correction proportional to their commutator.
Technically, for a pair of pure gauge fields ${\cal U}_i(x_\perp)$ and  ${\cal V}_i(x_\perp)$
we define ${\cal W}_i(x_\perp)={\cal U}_i(x_\perp)+{\cal V}_i(x_\perp)+gE_i(x_\perp;U,V)$ 
as a pure gauge field 
satisfying the equation $(i\partial_i+g[{\cal U}_i+{\cal V}_i,)E^i=0$. In the first order in $[U,V]$ 
this field has the form
\begin{eqnarray}
&&\hspace{-13mm}
E^a_i(U,V)~=~-(x_\perp|U{p^k\over p_\perp^2}U^\dagger
+V{p^k\over p_\perp^2}V^\dagger
\nonumber\\
&&\hspace{-3mm}
-~{p^k\over p_\perp^2}|^{ab}[{\cal U}_i,{\cal V}_k]^b
-i\leftrightarrow k)+O([U,V]^2)
\label{E1st}
\end{eqnarray}
where $[{\cal U}_i,{\cal V}_k]^a\equiv 2{\rm Tr}t^a[{\cal U}_i,{\cal V}_k]$.
The second, $[U,V]^2$, term of the expansion (\ref{E1st}) can be found in \cite{prd04} but we do not need it with our accuracy.

Throughout the paper, we use Schwinger notations for the propagator in the external field $(x|{1\over P^2}|y)$. For the  bare propagator it reduces to $(x|{1\over p^2}|y)$ and for the two-dimensional propagator in the transverse space we use the notation
$(x_\perp|{1\over p_\perp^2}|y_\perp)$ where $p_\perp^2=-p_ip^i$. Also,
$|f)$ denotes $\int\! d^2z_\perp f(z_\perp)|z_\perp)$ and later we will use the notation
$|0,f)\equiv\int\! d^2z_\perp f(z_\perp)|0,z_\perp)$.

The zero-order approximation  for the solution of the classical equations for the 
functional integral  
(\ref{mastegral}) can be taken as a superposition of pure gauge
fields in the forward, backward, left, and right quadrants of the space
(see Fig. \ref{fig:ansatz}):
\begin{figure}
\includegraphics[width=0.25\textwidth]{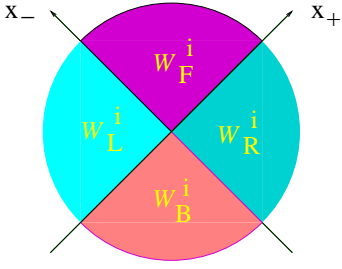}
\caption{Pure-gauge ansatz as a zero-order approximation to the classical field created by the collision of shock waves\label{fig:ansatz}.}
\end{figure}
%
\begin{eqnarray}
\bar{A}_\bullet^{(0)}&=&\bar{A}_\ast^{(0)}=0
\label{fild}\\
\bar{A}^{(0)i}&=&{\cal W}_F^i(x_\perp)\theta(x_\ast)\theta(x_\bullet)+
{\cal W}_L^i(x_\perp)\theta(-x_\ast)\theta(x_\bullet)
\nonumber\\
&+&{\cal W}_R^i(x_\perp)\theta(x_\ast)\theta(-x_\bullet)+
{\cal W}_B^i(x_\perp)\theta(-x_\ast)\theta(-x_\bullet)
\nonumber
\end{eqnarray}
where 
\begin{eqnarray}
&&\hspace{-3mm} 
{\cal W}_F^i={\cal U}_1^i+{\cal V}_1^i+E_F^i,~~~{\cal W}_L^i={\cal U}_2^i+{\cal V}_1^i+E_F^i
\nonumber\\
&&\hspace{-3mm}
{\cal W}_R^i={\cal U}_1^i+{\cal V}_2^i+E_F^i,~~~{\cal W}_B^i={\cal U}_2^i+{\cal V}_2^i+E_F^i
\end{eqnarray}
and $E_F^i(U_1,V_1)$, $E_L^i(U_2,V_1)$, $E_R^i(U_1,V_2)$, and 
$E_B^i(U_2,V_2)$ are given by Eq. (\ref{E1st}).
For the trial configuration (\ref{fild})
\begin{eqnarray}
&&\hspace{-3mm} 
\bar{F}_\bullet^{~ i}=\delta({2x_\ast\over s})
\{\theta(x_\bullet)({\cal W}_F^i-{\cal W}_L^i)+\theta(-x_\bullet)({\cal W}_R^i-{\cal W}_B^i)\}
  \nonumber\\
&&\hspace{-3mm}
\bar{F}_\ast^{~ i}=\delta({2x_\bullet\over s})
\{\theta(x_\ast)({\cal W}_F^i-{\cal W}_R^i)+\theta(-x_\ast)({\cal W}_L^i-{\cal W}_B^i)\}
 \nonumber\\
 \label{efs}
\end{eqnarray}
so
\begin{eqnarray}
D^i\bar{F}_{\bullet i}&=&\delta({2\over s}x_\ast)
\Big([\theta(x_\bullet)(\partial^i-i[\bar{A}^i,)({\cal W}_{Fi}-{\cal W}_{Li})
  \nonumber\\
&+&\theta(-x_\bullet)(\partial^i-i[\bar{A}^i,)({\cal W}_{Ri}-{\cal W}_{Bi})
  \nonumber\\
D^i\bar{F}_{\ast i}&=&\delta({2\over s}x_\bullet)
\Big(\theta(x_\ast)(\partial^i-i[\bar{A}^i,)({\cal W}_{Fi}-{\cal W}_{Ri})
  \nonumber\\
&+&\theta(-x_\ast)(\partial^i-i[\bar{A}^i,)({\cal W}_{Li}-{\cal W}_{Bi})
 \label{defs}
\end{eqnarray}
and 
\begin{eqnarray}
&&\hspace{-3mm} 
D_\ast\bar{F}_{\bullet i}~=~D_\ast\bar{F}_{\bullet i}~ 
\label{deefs}\\
&&\hspace{-3mm}=~\delta({2\over s}x_\ast)\delta({2\over s}x_\bullet)
({\cal W}_{Fi}-{\cal W}_{Ri}-{\cal W}_{Li}
+~{\cal W}_{Bi})
 \nonumber\\
&&\hspace{-3mm}=~\delta({2\over s}x_\ast)\delta({2\over s}x_\bullet)
(E_{Fi}-E_{Ri}-E_{Li}+E_{Bi})
 \nonumber
\end{eqnarray}

Next,  one
 shifts $A\rightarrow A+\bar{A}_i^{(0)}$  in the functional integral 
(\ref{mastegral})
and obtains
\begin{eqnarray}
&&\hspace{-3mm} 
e^{iS_{\rm eff}(V_1,V_2,U_1,U_2;\eta_1-\eta_2)}
~ \label{mastegral1}\\
&&\hspace{-3mm}=~\int\! DA~
 \exp\Big\{iS(\bar{A})+i\!\int\! d^4z(
 {1\over 2} A^\mu \bar{D}_{\mu\nu}A^\nu
 +T^\mu A_\mu)\Big\}.
 \nonumber
\end{eqnarray}
Here 
\begin{eqnarray}
&&\hspace{-3mm}
\bar{S}~=~{1\over 2}\!\int\!d^2z_\perp \Big\{({\cal V}_{1}-{\cal V}_{2})_i^a({\cal W}_F^i-{\cal W}_L^i+{\cal W}_R^i-{\cal W}_B^i)^{ia}
\nonumber \\
&&\hspace{-3mm}
~+({\cal U}_{1}-{\cal U}_{2})_i^a({\cal W}_F+{\cal W}_L-{\cal W}_R-{\cal W}_B)^{ia}-{1\over 2}({\cal W}_F
\nonumber \\
&&\hspace{-3mm}
~-{\cal W}_L+{\cal W}_R-{\cal W}_B)^{ia}({\cal W}_F+{\cal W}_L-{\cal W}_R-{\cal W}_B)_i^a
\label{klassikal}
\end{eqnarray}
is a sum of the action and source contributions due to the trial configuration (\ref{fild}),
$D_{\mu\nu}=D^2(\bar{A})g_{\mu\nu}-2i \bar{F}_{\mu\nu}$ 
is the inverse propagator in the background-Feynman gauge
\footnote{ Strictly speaking, the inverse propagator is the sum of  $D_{\mu\nu}$
and the second variational derivative of the source (\ref{generfun}), see  
 Ref.\cite{npb96}.}
 and $T_\mu$ is the linear term for our trial configuration:

\begin{eqnarray}
T_i&=&2\delta({2\over s}x_\ast)\delta(x_\bullet)({\cal W}_F^i-{\cal W}_L^i-{\cal W}_R^i+{\cal W}_B^i)
\nonumber \\
&=&2\delta({2\over s}x_\ast)\delta(x_\bullet)(E_F^i-E_L^i-E_R^i+E_B^i)
\nonumber \\
T_\ast&=&-{i\over 2}\delta({2\over s}x_\bullet)\Big(\theta(x_\ast)[{\cal V}_{1i}-{\cal V}_{2i},E_F^i+E_R^i]
\nonumber\\
&+&\theta(-x_\ast)[{\cal V}_{1i}-{\cal V}_{2i},E_L^i+E_B^i]\Big)
\nonumber\\
T_\bullet&=&-{i\over 2}\delta({2\over s}x_\ast)\Big(\theta(x_\bullet)[{\cal U}_{1i}-
{\cal U}_{2i},E_F^i+E_L^i]
\nonumber\\
& +&\theta(-x_\ast)[{\cal U}_{1i}-{\cal U}_{2i},E_R^i+E_B^i]\Big) 
\label{Ts}
\end{eqnarray}
The first line in this equation follows directly from Eq. (\ref{deefs}) while the two last 
lines are obtained by
adding Eqs. (\ref{defs}) and the corresponding first 
derivatives of the sources (\ref{cliqs})
\begin{eqnarray}
&&\hspace{-5mm}
\left.{\delta \over\delta A_\ast}\!\int\! d^2z_\perp ({\cal U}_1^{ai}-{\cal U}_2^{ai})_z
 (0,F_{\ast i},0)^a_z \right|_{A_\ast=0}
  \nonumber\\
&&\hspace{-5mm}=~
 \delta(x_\ast)
\Big\{[\theta(x_\bullet)(\partial^i-i[\bar{A}^i(\infty p_2+x_\perp),)({\cal U}_{1i}-{\cal U}_{2i})
  \nonumber\\
&&\hspace{-5mm}+~\theta(-x_\bullet)(\partial^i-i[\bar{A}^i(-\infty p_2+x_\perp),)({\cal U}_{1i}-{\cal U}_{2i})
\Big\}
  \nonumber\\
  &&\hspace{-5mm}
\left.{\delta \over\delta A_\bullet}
\!\int\! d^2z_\perp \Big[  ({\cal V}_1^{ai}-{\cal V}_2^{ai})_z
 [0,F_{\bullet i},0]^a_z\right|_{A_\bullet=0}
  \nonumber\\
&&\hspace{-5mm}=~\delta(x_\bullet)
\Big\{\theta(x_\ast)(\partial^i-i[\bar{A}^i(\infty p_1+x_\perp),)({\cal V}_{1i}-{\cal V}_{2i})
  \nonumber\\
&&\hspace{-5mm}+~\theta(-x_\ast)(\partial^i-i[\bar{A}^i(-\infty p_1+x_\perp),)({\cal V}_{1i}-{\cal V}_{2i})\}
 \label{fromsorses}
\end{eqnarray}
We get
\begin{eqnarray}
&&\hspace{-5mm}
\left.{\delta \over\delta A_\ast}\Big\{S_{\rm QCD}+\!\int\! d^2z_\perp ({\cal U}_1^{ai}-{\cal U}_2^{ai})_z
 (0,F_{\ast i},0)^a_z\Big\} \right|_{A_\ast=0}
  \nonumber\\
&&\hspace{-5mm}=~-
 \delta(x_\ast)
\Big\{\theta(x_\bullet)(\partial^i-i[\bar{A}^i(\infty p_2+x_\perp),)(E_{Fi}-E_{Li})
  \nonumber\\
&&\hspace{-5mm}+~\theta(-x_\bullet)
(\partial^i-i[\bar{A}^i(-\infty p_2+x_\perp),)(E_{Ri}-E_{Bi})
\Big\},
  \nonumber\\
  &&\hspace{-5mm}
\left.{\delta \over\delta A_\bullet}\Big\{S_{\rm QCD}+
\!\int\! d^2z_\perp \Big[  ({\cal V}_1^{ai}-{\cal V}_2^{ai})_z
 [0,F_{\bullet i},0]^a_z\Big\}\right|_{A_\bullet=0}
  \nonumber\\
&&\hspace{-5mm}=~-\delta(x_\bullet)
\Big\{\theta(x_\ast)(\partial^i-i[\bar{A}^i(\infty p_1+x_\perp),)(E_{Fi}-E_{Ri})
  \nonumber\\
&&\hspace{-5mm}+~
\theta(-x_\ast)(\partial^i-i[\bar{A}^i(-\infty p_1+x_\perp),)(E_{Li}-E_{Bi})\}
 \label{totalvklad}
\end{eqnarray}
Using  $\theta(0)={1\over 2}$ so that
 $\bar{A}^i(\infty p_2+x_\perp)={1\over 2}({\cal W}_{Ri}+{\cal W}_{Bi})$,
 $\bar{A}^i(-\infty p_2+x_\perp)={1\over 2}({\cal W}_{Li}+{\cal W}_{Bi})$,
 $\bar{A}^i(\infty p_1+x_\perp)={1\over 2}({\cal W}_{Fi}+{\cal W}_{Ri})$,
 $\bar{A}^i(-\infty p_1+x_\perp)={1\over 2}({\cal W}_{Li}+{\cal W}_{Bi})$,
 and the condition $(i\partial_i+[{\cal W}_i,)E^i=0$
one easily obtains Eq. (\ref{Ts}). 
\footnote{A careful analysis shows that the ``formula'' $\theta(0)={1\over 2}$ is not valid here.
It can be demonstrated that instead of 
${1\over 2}(E_F+E_L)\int_0^\infty dz_\ast A_\bullet(z_\ast)$ one should use
$E_F\int_0^\infty dz_\ast A^{(+)}_\bullet(z_\ast)+E_L\int_0^\infty dz_\ast A^{(-)}_\bullet(z_\ast)$ 
where $A^{(+)}$ and $A^{(-)}$ are the positive and negative
frequency parts of the field $A$. (With such $T$ one reproduces the correct set of fields $A_\ast$ and $A_\bullet$ given by Eq. (\ref{fiilds}) from the Appendix).  Fortunately, the corresponding contribution to the effective action is $\sim T_\ast T\bullet \sim [U,V]^3$ which exceeds our accuracy.}

Expansion in powers of $T$ in the functional integral (\ref{mastegral1}) 
yields the set of Feynman diagrams in the external fields (\ref{shoksum}) with the sources
(\ref{Ts}). The parameter of the expansion is $g^2[{\cal U}_i,{\cal V}_j]$ ($\sim [U,V]$, see Eq.
(\ref{defui})).

\section{\label{secteffect}The effective action}

\subsection{\label{secteffecta}The effective action in the lowest order}
The effective action (\ref{mastegral}) in the first nontrivial order in $[U,V]$ is 
given by the integration of linear terms (\ref{Ts}) with the Green functions
in the external field (\ref{fild})
\begin{eqnarray}
&&\hspace{0mm} 
iS_{\rm eff}(U,V)
~=~
\nonumber\\
&&\hspace{0mm}
-{1\over 2}\!\int\! d^4zd^4z'
T^a_\mu(z)\langle A^{\mu a}(z) A^{\nu b}(z')\rangle T^b_\nu(z')
 \label{seffgen}
\end{eqnarray}
It is easy to see that the term $\sim T_\ast T_\bullet$ is $\sim [U,V]^3$ so the leading
contribution $\sim [U,V]^2$ comes from the product of two $T_i$'s which has the form
\begin{equation}
\hspace{0mm} 
{i\over 2}\!\int\! d^2z_\perp d^2z'_\perp 
L^a_i(z_\perp)(0,z_\perp|{1\over P^2+i\epsilon}|0, z'_\perp)^{ab} L^{bi}(z'_\perp)
 \label{seffgen1}
\end{equation}
where 
\begin{eqnarray}
\hspace{-3mm} 
L_i&\equiv& 2(E_F^i-E_L^i-E_R^i+E_B^i)
\nonumber\\
&=&2({\cal W}_F^i-{\cal W}_L^i-{\cal W}_R^i+{\cal W}_B^i)
\label{lvertex}
\end{eqnarray}
is actually the transverse part of the Lipatov vertex of the gluon emission by the
scattering of two shock waves in the first order in $[U,V]$ (see Appendix).
As we shall see below, the main logarithmic contribution to the integral (\ref{seffgen1})
comes from the region $z_\perp\rightarrow z'_\perp$ where one can replace the propagator
in the background field by the bare propagator. One obtains
\begin{eqnarray}
&&\hspace{-3mm} 
ig^2{s\over 2}\!\int\!{d\alpha d\beta\over 8\pi^2}
(0,L_i^a|{1\over \alpha\beta s-p^2_\perp+i\epsilon}
|0,L^{ia})
\label{effgen1a}
\end{eqnarray}

The integral (\ref{effgen1a}) is formally divegrent. 
Within the LLA accuracy, one can cut the integration off at the width of the shock waves
$\lambda\sim \sqrt{s\over m^2}e^{-\eta_1/2}$, 
$\rho\sim \sqrt{s\over m^2}e^{\eta_2/2}$ and obtain:
\begin{eqnarray}
&&\hspace{-3mm} 
ig^2{s\over 2}\!\!\int\!{d\alpha d\beta\over 8\pi^2}
e^{-i(\alpha\lambda+\beta\rho)}
(0,L_i^a|{1\over \alpha\beta s-p^2_\perp+i\epsilon}
|0,L^{ia})
\nonumber\\
&&\hspace{0mm}
=~{\alpha_s\Delta\eta\over 4}\!\int d^2z_\perp L_i^a(z_\perp)L^{ia}(z_\perp)
\label{withslope}
\end{eqnarray}
where $\Delta\eta=\eta_1-\eta_2$ is our rapidity interval.

In addition,  within the LLA approximation the zero-order term (\ref{klassikal}) can be simplified to
\begin{eqnarray}
\bar{S}&=&\int\!d^2z_\perp({\cal V}_{1}-{\cal V}_{2})^{ia}({\cal U}_{1}-{\cal U}_{2})_i^a
\label{klassi}
\end{eqnarray}
Indeed, it is easy to see that in the r.h.s. of Eq. (\ref{klassikal}) the terms $\sim [U,V]$ 
cancel while the the terms $\sim [U,V]^2$ are not multiplied by $\Delta\eta$ so they
can be omitted in the LLA  (the  Eq. (\ref{withslope}) is $\sim [U,V]^2 \Delta\eta$).
It is worth noting that Eq. (\ref{klassi}) is the usual light-cone lattice action
\cite{pirner} in the limit when transverse size of the plaquette vanishes and the longitudinal
increases to infinity.
Thus, the effective action in the first order can  be represented as
\begin{eqnarray}
&&\hspace{0mm} 
S_{\rm eff}(U,V)~=~\!\int\!d^2z_\perp 
\Big\{({\cal V}_{1}-{\cal V}_{2})_i^a({\cal U}_{1}-{\cal U}_{2})_i^a\nonumber\\
&&\hspace{0mm}-~i
{\alpha_s\Delta\eta\over 4}\!\int d^2z_\perp L_i^a(z_\perp)L^{ia}(z_\perp)\Big\}
\label{effact}
\end{eqnarray}
We shall see below that $L_i$ is the Lipatov vertex of the gluon emission by the
scattering of two shock waves in the first order in $[U,V]$. Note that $S_{\rm eff}$ given by Eq. (\ref{effact}) is
invariant with respect to rotation of the sources 
\begin{equation}
U_j\rightarrow U_j\Omega,~~~~ 
V_j\rightarrow V_j\Omega
\label{rotat}
\end{equation}
For the first term in the r.h.s. of Eq. (\ref{effact}) it is trivial while for the second
it follows from the gauge-invariant form discussed in the next section, see Eq. (\ref{rotati}).

For future applications we will rewrite the effective action
(\ref{effact}) as a Gaussian integration over the auxiliary field 
$\lambda$ coupled to Lipatov vertex (\ref{els}):
\begin{eqnarray}
&&\hspace{0mm} 
e^{iS_{\rm eff}(U,V)}
~=~e^{i\int d^2z_\perp~({\cal V}_{1}-{\cal V}_{2})^{ia}({\cal U}_{1}-{\cal U}_{2})_i^a}
\label{effacta}\\
&&\hspace{0mm}\times~\int D\lambda 
\exp\Big\{-\alpha_s\Delta\eta\int\! d^2z_\perp
(\lambda^a_i\lambda^{ai}
-L^a_i\lambda^{ai})\Big\}
\nonumber
\end{eqnarray}
%

\subsection{\label{secteffectb}Nonlinear evolution equation from the effective action}

Let us prove now that the effective action (\ref{effacta}) agrees with the 
non-linear evolution equation. 
 To find the evolution of the dipole $U_xU^\dagger_y$, we need to consider
the effective action for the weak source $V$. From eq. (\ref{E1st}) one sees that
at small $g{\cal V}_i\sim\partial_iV$ 
\begin{equation}
\hspace{-0mm} 
L^a_i(x_\perp)=
-2(x|U_1^\dagger{p_ip^k\over p_\perp^2}U_1-U_2^\dagger{p_ip^k\over p_\perp^2}U_2|^{ab}
({\cal V}_1-{\cal V}_2)^b_k)
\label{lsmol}
\end{equation}
and Eq. (\ref{mastegral}) can be rewritten as
\begin{eqnarray}
&&\hspace{-3mm} 
\int\!\! DA
 \exp\Big\{iS(A) +
i \!\int\! d^2z_\perp \Big[  ({\cal V}_1^{ai}-{\cal V}_2^{ai})_z
 [0,F_{\bullet i},0]^a_z
\nonumber\\
&&\hspace{-3mm} +~
({\cal U}_1^{ai}-{\cal U}_2^{ai})_z
 (0,F_{\ast i},0)^a_z
 \Big] \Big\}
 \nonumber\\
&&\hspace{-5mm} 
=\int D\lambda 
e^{\int\! d^2z_\perp\{-\alpha_s\Delta\eta\lambda^a_i\lambda^{ai} 
+i({\cal V}_1-{\cal V}_2)_i^a(\tilde{{\cal U}}_1^i-\tilde{{\cal U}}_2^i)^a\}}.
  \label{effactu}
\end{eqnarray}
Here
\begin{eqnarray}
\hspace{-5mm}
\tilde{U}_1&=&e^{2\alpha_s\Delta\eta{\partial_i\over \partial_\perp^2}
(U_1\lambda^i U_1^\dagger)}U_1
,\\
\tilde{U}_2&=&e^{2\alpha_s\Delta\eta{\partial_i\over \partial_\perp^2}
(U_2\lambda^i U_2^\dagger)}U_2
\label{tildeus}
\end{eqnarray}
so that
\begin{eqnarray}
\tilde{{\cal U}}_{1i}^a~&=&~{\cal U}_{1i}^a+
2\alpha_s\Delta\eta(U_1^\dagger{\partial_i\partial^k\over\partial_\perp^2}U_1)^{ab}\lambda^b_k\\
 \tilde{{\cal U}}_{2i}^a~&=&~{\cal U}_{2i}^a+
2\alpha_s\Delta\eta(U_2^\dagger{\partial_i\partial^k\over\partial_\perp^2}U_2)^{ab}\lambda^b_k
\label{tildeus1}
\end{eqnarray}
where we need only the first term in expansion in $\lambda_i$ in $\lambda$. 
\footnote{To cancel the UV divergence in the gluon-reggeization term
 $\sim 2t^aU_xt^b(x|{p_i\over p_\perp^2}U^{ab}{p_i\over p_\perp^2}|y)$ we need the second-order contribution
 $c_F(x|{1\over p_\perp^2}|x)U_x+c_F(x|{1\over p_\perp^2}|x)U_y$. 
 However, since the pure divergency
 is set to zero in the dimensional regularization, at least within this regularization the first term is sufficient.}

 To find the
evolution of the color dipole (\ref{dipole}) we should expand Eq. 
(\ref{effactu}) in powers of ${\cal V}_{1i}-{\cal V}_{2i}$ and use the 
formula 
\begin{eqnarray}
&&\hspace{0mm}
[0,\infty p_1]_x[x_\perp+\infty p_1,y_\perp+\infty p_1][\infty p_1,0]_y^{\eta_2}
\nonumber\\
&&\hspace{0mm}=~Pe^{ig\!\int_x^y\!dz_i[0,F_{\bullet i},0]_z}
\end{eqnarray}
which results in
\begin{eqnarray}
&&\hspace{0mm}
U^\dagger_{1y}U_{1x}U_{2x}^\dagger U_{2y}
\nonumber\\
&&\hspace{0mm}~=\int D\lambda ~
e^{-\alpha_s\Delta\eta\int\! d^2z_\perp\lambda^a_i\lambda^{ai}}~
\tilde{U}_{1y}^\dagger\tilde{U}_{1x}\tilde{U}_{2x}^\dagger \tilde{U}_{2y}
\end{eqnarray}
Performing the Gaussian integration over $\lambda$ one obtains after some algebra
\begin{eqnarray}
&&\hspace{-4mm}
{\rm tr}\{U_{1x}U_{2x}^\dagger U_{2y}U^\dagger_{1y}\}
\label{bk}\\
&&\hspace{-4mm}=~
{\rm tr}\{U_{1x}U_{2x}^\dagger U_{2y}U^\dagger_{1y}\}
+{\alpha_s\Delta\eta\over 4\pi^2}\!\int\! d^2z_\perp 
{(x-y)_\perp^2\over(x-z)_\perp^2(z-y)_\perp^2}
\nonumber\\
&&\hspace{-4mm}
\times~({\rm tr}\{U_{1x}U_{2x}^\dagger U_{2z}U^\dagger_{1z}\}
{\rm tr}\{U_{1z}U_{2z}^\dagger U_{2y}U^\dagger_{1y}\}
\nonumber\\
&&\hspace{40mm}
-N_c{\rm tr}\{U_{1x}U_{2x}^\dagger U_{2y}U^\dagger_{1y}\})
\nonumber
\end{eqnarray}
which is the non-linear evolution equation \cite{npb96,yura}
  for the Wilson-line operator 
$U_x=U_{1x}U_{2x}^\dagger =[\infty e,-\infty e]_x$.

\subsection{Gauge-invariant representation of the  first-order effective action $L_iL^i$}
\begin{figure}
\includegraphics[width=0.27\textwidth]{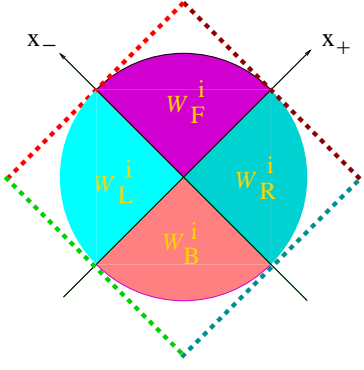}
\caption{Gauge-invariant form of the effective action\label{fig:diamond1}.}
\end{figure}

Our expression for the $\Delta\eta$ term in the effective action, proportional to 
the square of the Lipatov vertex $({\cal W}_F^i-{\cal W}_L^i-{\cal W}_R^i+{\cal W}_B^i)^a$, 
 was obtained in the axial-type gauges. 
It can be rewritten it in the gauge-invariant  ``diamond''
form of trace of four Wilson lines at $x_{\bullet,\ast}=\pm\infty$
(see Fig. \ref{fig:diamond1}) as suggested in a recent paper \cite{smith}.
The ``diamond" Wilson loop is defined as follows
\begin{eqnarray}
&&\hspace{-4mm}
\Diamond(x_\perp)\equiv~
 {\rm tr}\{[-\infty p_1,F_{\bullet i},\infty p_1]_{\infty p_2}
[\infty p_2,F_{\ast i},-\infty p_2]_{\infty p_1}
\nonumber\\
&&\hspace{3mm}
\times~[\infty p_1,-\infty p_1]_{-\infty p_2}[-\infty p_2,\infty p_2]_{-\infty p_1}
\nonumber\\
&&\hspace{3mm}
+~{\rm tr}[-\infty p_1,\infty p_1]_{\infty p_2}[\infty p_2,F_{\ast i},-\infty p_2]_{\infty p_1}
\nonumber\\
&&\hspace{3mm}
\times~[\infty p_1,F_{\bullet i},-\infty p_1]_{-\infty p_2}[-\infty p_2,\infty p_2]_{-\infty p_1}
\nonumber\\
&&\hspace{3mm}
+~ {\rm tr}[-\infty p_1,\infty p_1]_{\infty p_2}[\infty p_2,-\infty p_2]_{\infty p_1}
\nonumber\\
&&\hspace{3mm}
\times~[\infty p_1,F_{\bullet i},\-\infty p_1]_{-\infty p_2}
[-\infty p_2,F_{\ast i},\infty p_2]_{-\infty p_1}
\nonumber\\
&&\hspace{3mm}
+~ {\rm tr}[-\infty p_1,F_{\bullet i},\infty p_1]_{\infty p_2}
[\infty p_2,-\infty p_2]_{\infty p_1}
\nonumber\\
&&\hspace{3mm}
\times~[\infty p_1,-\infty p_1]_{-\infty p_2}
[-\infty p_2,F_{\ast i},\infty p_2]_{-\infty p_1}\}
\label{dia1}
\end{eqnarray}
where 
the transverse arguments in all  Wilson lines are $x_\perp$. Next, define this
``diamond''  as a function of the sources
\begin{eqnarray}
&&\hspace{-3mm}
\Diamond(U_1,U_2,V_1,V_2)~\equiv~
\label{dia2}\\
&&\hspace{-3mm}
{\cal N}^{-1}\!\int\!\! DA~\Diamond(A)~
 \exp\Big(iS(A) +
i \!\int\! d^2z_\perp \Big\{  ({\cal V}_1^{ai}-{\cal V}_2^{ai})_z
\nonumber\\
&&\hspace{13mm}
\times~
 [0,F_{e_1 i},0]^a_z +
({\cal U}_1^{ai}-{\cal U}_2^{ai})_z
 (0,F_{e_2 i},0)^a_z
\Big\} \Big)
\nonumber
\end{eqnarray}
(In the Appendix we demonstrate that the trace of four Wilson lines  
\begin{eqnarray}
&&\hspace{-3mm}
 {\rm tr}\{[-\infty e_1,\infty e_1]_{\infty e_2}
[\infty e_2,-\infty e_2]_{\infty e_1}
\nonumber\\
&&\hspace{-3mm}\times~
[\infty e_1,-\infty e_1]_{-\infty e_2}[-\infty e_2,\infty e_2]_{-\infty e_1}\}~=~1
\end{eqnarray}
is trivial in the leading order.)

Note that $\Diamond(U,V)$ is invariant with respect to the rotation of all sources by  one
gauge matrix
$\Omega(x_\perp)$
\begin{eqnarray}
\Diamond(U_1\Omega,U_2\Omega, V_1\Omega,V_2\Omega)=\Diamond(U_1,U_2, V_1,V_2)
\label{rotati}
\end{eqnarray}
since it can be absorbed by gauge transformation of the fields 
$A_\mu\rightarrow \Omega^\dagger A_\mu\Omega+{i\over g}\Omega^\dagger \partial_\mu\Omega$
in the functional integral Eq. (\ref{dia2}).

Now we can prove that  the square of Lipatov vertex can be expressed as the ``diamond'' Wilson loop:
\begin{eqnarray}
&&\hspace{-3mm}
{1\over 4}L^a_i(U,V)L^{ai}(U,V)~=~\Diamond(U,V)
\label{diamond}
\end{eqnarray}
Indeed,  it is easy to see that for the trial 
configuration (\ref{fild}) the Eq. (\ref{dia1})
reduces to 
\begin{eqnarray}
&&\hspace{-3mm}
({\cal W}_{Fi}-{\cal W}_{Li})^a({\cal W}_F^i-{\cal W}_R^i)^a+({\cal W}_{Ri}-{\cal W}_{Fi})^a
\nonumber\\
&&\hspace{-3mm}
\times~({\cal W}_R^i-{\cal W}_B^i)^a
+({\cal W}_{Bi}-{\cal W}_{Ri})^a({\cal W}_B^i-{\cal W}_L^i)^a
\nonumber\\
&&\hspace{-3mm}
+~({\cal W}_{Li}-{\cal W}_{Bi})^a({\cal W}_L^i-{\cal W}_F^i)^a~=
\nonumber\\
&&\hspace{-3mm}
({\cal W}_{Fi}-{\cal W}_{Li}-{\cal W}_R^i+{\cal W}_{Bi})^a
({\cal W}_F^i-{\cal W}_L^i-{\cal W}_R^I+{\cal W}_B^i)^a
\nonumber
\end{eqnarray}
 which coincide with 
the l.h.s. of the Eq.  (\ref{diamond}). 
In the covariant-type gauges where $A_i\rightarrow 0$ as $x_\parallel\rightarrow \infty$
the r.h.s of the Eq. (\ref{dia1}) can be rewritten as 
\begin{eqnarray}
&&\hspace{-1mm}
\partial_i M_1\partial^i M_2 M_3^\dagger M_4^\dagger
+M_1\partial_i M_2 \partial^i M_3^\dagger M_4^\dagger
\nonumber\\
&&\hspace{-1mm}
M_1 M_2\partial_i  M_3^\dagger\partial_i  M_4^\dagger
+\partial_i M_1 M_2 M_3^\dagger\partial_i M_4^\dagger
\label{diamonda}
\end{eqnarray}
where $M_1=[-\infty p_1,\infty p_1]_{\infty p_2}$, 
 $M_2=[\infty p_2,-\infty p_2]_{\infty p_1}$, 
 $M_3= [\infty p_1,-\infty p_1]_{-\infty p_2}$, and  
 $M_4=[-\infty p_2,\infty p_2]_{-\infty p_1}$.
 Eq. (\ref{diamonda}) is the expression obtained recently in \cite{smith} in the framework 
 of the Hamiltonian approach
 (see also \cite{kl05,mu05}). The corresponding form of our effective action is the following:
\begin{eqnarray}
&&\hspace{-3mm}
{1\over 4}L^a_i(U,V)L^{ai}(U,V)~
\nonumber\\
&&\hspace{-1mm}
=~\partial_i(W_L W^\dagger_F)\partial^i(W_F W^\dagger_R)
W_R W^\dagger_BW_B W^\dagger_L
\nonumber\\
&&\hspace{-1mm}
+~W_L W^\dagger_F\partial^i(W_F W^\dagger_R)
\partial_i(W_R W^\dagger_B)W_B W^\dagger_L
\nonumber\\
&&\hspace{-1mm}
+~W_L W^\dagger_FW_F W^\dagger_R
\partial_i(W_R W^\dagger_B)\partial^i(W_B W^\dagger_L)
\nonumber\\
&&\hspace{-1mm}
+~\partial_i(W_L W^\dagger_F)W_F W^\dagger_R
W_R W^\dagger_B\partial^i(W_B W^\dagger_L)
\label{moidiamond}
\end{eqnarray}

The Eq. (\ref{diamond})  links the representation in terms of the effective degrees of freedom (Wilson lines in our case)
 with the representation in terms of gluons of the underlying Yang-Mills theory via Eq. (\ref{dia2}).
The  remarkable feature of the gauge-invariant form (\ref{diamond}) 
is its universality - 
 if one writes the effective action in terms of some other degrees of
 freedom (say, reggeized gluons \cite{leffaction}) one should recover 
 Eq. (\ref{diamond}) once these new effective degrees of freedom are expressed in terms of gluons.

\section{Functional integral over the dynamical Wilson lines}
 

\subsection{Effective action as the integral over the Wilson lines} 

\begin{figure}
\includegraphics[width=0.45\textwidth]{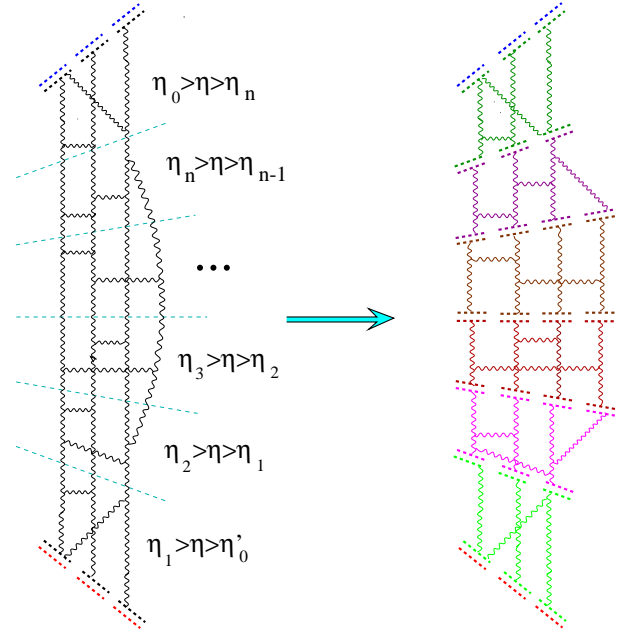}
\caption{Effective action factorized in $n$ functional integrals\label{fig:figentimes}.}
\end{figure}

In this section we will rewrite the functional integral for the 
effective
action (\ref{mastegral}) in terms of Wilson-line variables. To this end, 
let us use
the factorization formula (\ref{faktor}) $n$ times as shown in 
Fig.~\ref{fig:figentimes}.
The effective action factorizes then into a product 
of $n$ independent functional integrals over the gluon fields labeled
by index $k$:
%
\begin{eqnarray}
&&\hspace{-2mm}e^{iS_{\rm eff}(U,V;\eta)}=\label{pokabez}\\
&&\hspace{-2mm}\int\! DA^1DA^2\ldots DA^{n+1}
\exp i\Big\{({\cal V}_1^i-{\cal V}_2^i)({\cal U}^{n+1}_{1i}-{\cal U}^{n+1}_{2i})\nonumber\\
&&\hspace{-2mm}+~
S(A_{n+1})+~
({\cal V}_1^{n+1,i}-{\cal V}_2^{n+1,i})({\cal U}^{ni}_1-{\cal U}^{ni}_2)\nonumber\\
&&\hspace{-2mm}
+~S(A_n)+\ldots +({\cal V}_1^{2i}-{\cal V}_2^{2i})({\cal U}_{1i}^1-{\cal U}_{2i}^1)+S(A_1)\nonumber\\
&&\hspace{22mm}
+~({\cal V}_1^{1i}-{\cal V}_2^{1i})({\cal U}_{1i}-{\cal U}_{1i})\Big\} ~,
\nonumber
\end{eqnarray}
where the integrals over $x_{\perp}$ and summation over the color indices are
implied.  As usually, ${\cal U}^{k,i}_j={i\over g}U^{\dagger k}_j\partial^iU^k_j$ and 
${\cal V}^{k,i}_j={i\over g}V^{\dagger k}_j\partial^iV_j^k$
where
\begin{eqnarray}
U^k_{1(2)}(x_{\perp})&=&Pe^{ig\int_0^{\pm\infty}du \ n_k^{\mu}
A_{k,\mu}(un^k+x_{\perp})} ,\nonumber\\
V^k_{1(2)}(x_{\perp})&=&Pe^{ig\int_0^{\pm\infty}du \ n_{k-1}^{\mu}
A_{k,\mu}(un^{k-1}+x_{\perp})}
\label{nov2}
\end{eqnarray}
and the vectors $n_k$ are ordered in rapidity: 
$\eta_0>\eta_n>\eta_{n-1}\ldots\eta_2>\eta_1>\eta'_0$.
To disentangle integrations over different $A^k$, we rewrite 
$\exp[i({\cal V}_1^{k+1,i}-{\cal V}_2^{k+1,i})({\cal U}^{k,i}_1-{\cal U}^{k,i}_2)]$ at each
``rapidity divide'' $\eta_k$
as an integral over the auxiliary group variables $\hat{V}^{k+1}_{1,2}$ and $\hat{U}^k_{1,2}$ using the formula 
%
\begin{eqnarray}
&&\hspace{-2mm}
e^{i\int dx_{\perp}V_iU^i}~=~
\det (\partial_i-ig{\cal V}_i)(\partial^i-ig{\cal U}^i)\label{nov3}\\
&&\hspace{-2mm}
\times
\int\! D\hat{V}(x_{\perp})D\hat{U}(x_{\perp})e^{i\int dx_{\perp}{\cal V}_i\hat{\cal U}^i+
i\int\! dx_{\perp}\hat{\cal V}_i{\cal U}^i-i\int dx_{\perp}\hat{\cal V}_i\hat{\cal U}^i}.
\nonumber
\end{eqnarray}
(where $\hat{\cal V}_i\equiv \hat{V}^\dagger {i\over g}\partial_i\hat{V}$ and
 $\hat{\cal U}_i\equiv \hat{U}^\dagger {i\over g}\partial_i\hat{U}$).
The determinant gives the perturbative non-logarithmic corrections 
of the same order as the corrections to the factorization 
formula (\ref{faktor}). In the LLA they can be ignored, 
consequently, we obtain
\begin{eqnarray}
&&\hspace{-3mm}
e^{iS_{\rm eff}(U_1,U_2,V_1,V_2,\eta_1-\eta_2)}~
\label{nov4}\\
&&\hspace{-3mm}=~
\int\!\Pi_{k=0}^{n+1} DA^k~\Pi_{k=0}^{n}  D\hat{U}_1^kD\hat{U}_2^k
D\hat{V}_1^kD\hat{V}_2^k 
\nonumber\\
&&\hspace{-3mm}\times~
\exp i\Big\{ ({\cal V}_1^i-{\cal V}_2^i)({\cal U}^{n+1}_{1i}-{\cal U}^{n+1}_{1i})
+S(A^{n+1})+({\cal V}_{1i}^{n+1}
\nonumber\\
&&\hspace{-3mm}
-~{\cal V}_{2i}^{n+1})(\hat{\cal U}_1^{ni}-\hat{\cal U}_2^{ni})
-(\hat{\cal V}^n_{1i}-\hat{\cal V}^n_{2i})(\hat{\cal U}^{n,i}_1-\hat{\cal U}^{n,i}_2)
+\ldots
\nonumber\\
&&\hspace{-3mm}
+~({\cal V}_{1i}^3-{\cal V}_{2i}^3)(\hat{\cal U}^{2i}_1-\hat{\cal U}^{2i}_2)-(\hat{\cal V}_1^{2i}-\hat{\cal V}_2^{2i})(\hat{\cal U}^2_{1i}-\hat{\cal U}^2_{2i})
\nonumber\\
&&\hspace{-3mm}
+~(\hat{\cal V}^2_{2i}-\hat{\cal V}^2_{1i})({\cal U}^{2i}_1-{\cal U}^{2i}_2)+
S(A^2)+
({\cal V}^2_{1i}-{\cal V}^2_{2i})(\hat{\cal U}_1^{1i}
\nonumber\\
&&\hspace{-3mm}-~\hat{\cal U}_2^{1i})-
(\hat{\cal V}^1_{1i}-\hat{\cal V}^1_{2i})(\hat{\cal U}_1^{1i}-\hat{\cal U}_2^{1i})
+(\hat{\cal V}^1_{1i}-\hat{\cal V}^1_{2i})
\nonumber\\
&&\hspace{3mm}\times~({\cal U}^{1i}_1-{\cal U}^{1i}_2)+S(A^1)+
({\cal V}_1^{1i}-{\cal V}_2^{1i})(U_{1i}-U_{2i})\Big\} .
\nonumber
\end{eqnarray}
Now we can integrate over the gluon fields $A_k$. Using the results 
of the previous Section, we get 
\begin{eqnarray}
&&\hspace{-2mm}
\int\! DA_k \exp\{i(\hat{\cal V}^k_{1i}-\hat{\cal V}^k_{2i})({\cal U}_1^{k,i}-{\cal U}_2^{k,i})
+iS(A_k)
\nonumber\\
&&\hspace{12mm}+~i({\cal V}^k_{1i}-{\cal V}^k_{2i})
(\hat{\cal U}^{k-1,i}_1-\hat{\cal U}^{k-1,i}_2)\}
\nonumber\\
&&\hspace{22mm}=~
e^{iS_{\rm eff}(\hat{V}_1^k,\hat{V}_2^k,\hat{U}_1^{k-1},\hat{U}_2^{k-1};\Delta\eta)}
\label{nov5}
\end{eqnarray}
where at sufficiently small $\Delta\eta$
\begin{eqnarray}
&&\hspace{-2mm}
S_{\rm eff}(\hat{V}_1^k,\hat{V}_2^k,\hat{U}_1^{k-1},\hat{U}_2^{k-1};\Delta\eta)
~=~
(\hat{\cal V}^k_{1i}-\hat{\cal V}^k_{2i})(\hat{\cal U}^{k-1,i}_1
\nonumber\\
&&\hspace{-2mm}
-~\hat{\cal U}^{k-1,i}_2)
~-~i{\alpha_s\Delta\eta\over 4}L_i(\hat{\cal V}^k,\hat{\cal U}_{k-1})
L^i(\hat{\cal V}^k,\hat{\cal U}_{k-1})
\label{nov6}
\end{eqnarray}
Performing the integrations over $A^k$ we get
\begin{eqnarray}
&&\hspace{-2mm}e^{iS_{\rm eff}(U,V,\eta_1-\eta_2)}~=~\int 
\Pi_{k=0}^{n}  D\hat{U}_1^kD\hat{U}_2^kD\hat{V}_1^kD\hat{V}_2^k
\nonumber\\
&&\hspace{-2mm}\times~
\exp i\Big\{ ({\cal V}_1^i-{\cal V}_2^i)(\hat{\cal U}_{1i}^n-\hat{\cal U}_{2i}^n)
-{i\alpha_s\over 4} L^2(V,\hat{U}^n)\Delta\eta
\nonumber\\
&&\hspace{-2mm}-~(\hat{\cal V}^n_{1i}-\hat{\cal V}^n_{2i})(\hat{\cal U}^{n,i}_1-\hat{\cal U}^{n,i}_2)+\ldots
\nonumber\\
&&\hspace{-2mm}
-~(\hat{\cal V}_1^{2i}-\hat{\cal V}_2^{2i})(\hat{\cal U}^2_{1i}-\hat{\cal U}^2_{2i})
+(\hat{\cal V}^2_{1i}-\hat{\cal V}^2_{2i})(\hat{\cal U}_1^{1i}-\hat{\cal U}_2^{1i})
\nonumber\\
&&\hspace{-2mm}
-~{i\alpha_s\over 4}L^2(\hat{V}^2,\hat{U}^1)\Delta\eta
-(\hat{\cal V}^1_{1i}-\hat{\cal V}^1_{2i})(\hat{\cal U}_1^{1i}-\hat{\cal U}_2^{1i})
\nonumber\\
&&\hspace{-2mm}
+~(\hat{\cal V}^1_{1i}-\hat{\cal V}^1_{2i})(U_{1i}-U_{2i})
-{i\alpha_s\over 4}L^2(\hat{V}^1,U)\Delta\eta
\Big\} .
\end{eqnarray}
In the continuum limit $n\rightarrow\infty$ we obtain the following functional
integral for the effective action
\begin{eqnarray}
&&\hspace{-4mm}
e^{iS_{\rm eff}(U_1(x),U_2(x),V_1(x),V_2(x);\eta_1-\eta_2)}
 \label{mastegrac}\\
&&\hspace{-4mm}
=~\left.\int \!\Pi_{j=1,2}DV_j(x,\eta)DU_j(x,\eta)\right|_{U_j(x,\eta_2)=U_j(x)}
\nonumber\\
&&\hspace{-4mm}\times~\exp\Bigg[\int\!d^2x
\Big(i[{\cal V}^a_{1i}(x)-{\cal V}^a_{2i}(x)]
[{\cal U}_1^{ai}
(x,\eta)-{\cal U}_2^{ai}(x,\eta)]
\nonumber\\
&&\hspace{-4mm}+~ \int_{\eta'_0}^{\eta_0}\!\!d\eta
\Big\{-i[{\cal V}^a_{1i}(x,\eta)-{\cal V}^a_{2i}(x,\eta)]
\nonumber\\
&&\hspace{32mm}\times~{\partial\over\partial\eta}[{\cal U}_1^{ai}(x,\eta)-{\cal U}_2^{ai}(x,\eta)]
\nonumber\\
&&\hspace{-4mm}
+~ {\alpha_s\over 4}L^a_i(V(x,\eta),U(x,\eta ))
L^{ai}(V(x,\eta),U(x,\eta))\Big\}\Big) \Bigg]
\nonumber
\end{eqnarray}
where we displayed the color indices explicitly and removed the hat
from the notation of the integration variables.
This looks like the functional integral over the canonical coordinates $U$ and 
canonical momenta $V$ with the (non-local) Hamiltonian $L^2(V,U)$.
The rapidity $\eta$ serves as the time variable for
this system. The above representation of the effective action as an integral over the dynamical Wilson-line variables is the main result of this paper.

Note that the $L_iL^i$ term in the exponent in  (\ref{mastegrac}) is invariant under the rotations (\ref{redun})
\begin{equation}
U_j(x,\eta)\rightarrow U_j(x,\eta)\Omega(x,\eta),~~~ V_j(x,\eta)\rightarrow V_j(x,\eta)\Omega(x,\eta)
\label{rotation}
\end{equation}
(see. Eq. (\ref{rotati})), but the  term 
  $\sim {\cal U}{\partial\over\partial\eta}{\cal V}$ preserves only the $\eta$-independent symmetry
\begin{equation}
U_j(x,\eta)\rightarrow U_j(x,\eta)\Omega(x), ~~~V_j(x,\eta)\rightarrow V_j(x,\eta)\Omega(x).
\label{rotatio}
\end{equation}
This probably means that the term $\sim {\cal U}{\partial\over\partial\eta}{\cal V}$ should be adjusted
by a  $\sim[U,V]^2$ correction (not important in the LLA) so that the full symmetry (\ref{rotation}) is restored.

The idea how to use the factorization formula to rewrite the functional integral in terms of Wilson lines was formulated in Ref. \cite{mobzor} where the first-order  effective action was obtained (the expression in terms of square
of Lipatov vertex is given in \cite{prd04}). However, the additional redundant gauge symmetry (\ref{redun}) was fixed by the requirement that there is no field at
 $t\rightarrow -\infty$ which correspond to the choice $U_2=0$ and $V_2=0$ for the two 
 colliding shock waves. In this case, one obtains 
the  functional integral in terms of only two variables, $U$ and $V$, at a price of a more complicated form of the effective action \cite{mobzor}.

\begin{figure}
\includegraphics[width=0.18\textwidth]{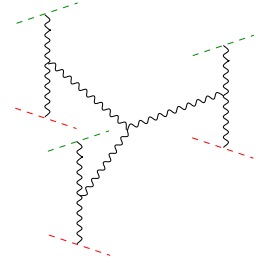}
\caption{Feynman diagram for a typical $[U,V]^3$ term\label{fig:3regge}.}
\end{figure}

 It should be noted that $L_i^2(U,V)$ is only the first term
of the expansion of the true high-energy effective action $K(U,V)$ in powers 
of $[U,V]$. An example of the next-order, $\sim[U,V]^3$, contribution to $K(V,U)$
which is missing in the effective action (\ref{mastegrac}) is presented in Ref \onlinecite{mobzor}, 
see Fig. \ref{fig:3regge}.

\subsection{Functional integral for the non-linear evolution}
It is instructive do demonstrate  that the functional integral (\ref{mastegrac})
reproduces the non-linear evolution in the case of one small source. Basically, we recast the arguments of the Sect. \ref{secteffectb} in the language of functional integrals. 

First, note that at small $V$ the functional integral over ${\cal V}$ is Gaussian
(see the Eq. (\ref{lsmol})). It is convenient to introduce the  ``gaussian noise'' 
associated with the Lipatov vertex and rewrite the functional 
integral (\ref{mastegrac}) as:
\begin{eqnarray}
&&\hspace{-4mm}
e^{iS_{\rm eff}(U_1(x),U_2(x),V_1(x),V_2(x);\eta_1-\eta_2)} \label{mastegram}\\
&&\hspace{-4mm}
=\left.\int \!\Pi_{j=1,2}DV_j(x,\eta)DU_j(x,\eta)\right|_{U_j(x,\eta_2)=U_j(x)}
\!\!D\lambda^a_i(x,\eta)
\nonumber\\
&&\hspace{-4mm}\times~\exp\int\!d^2x
\Big(i[{\cal V}^a_{1i}(x)-{\cal V}^a_{2i}(x)]
[{\cal U}_1^{ai}
(x,\eta)-{\cal U}_2^{ai}(x,\eta)]
\nonumber\\
&&\hspace{-4mm}+~ \int_{\eta'_0}^{\eta_0}\!\!d\eta
\Big\{-\alpha_s\lambda^a_i(x,\eta)\lambda^{ai}(x,\eta)\nonumber\\
&&\hspace{-4mm}
-~i[{\cal V}^a_{1i}(x,\eta)-{\cal V}^a_{2i}(x,\eta)]
{\partial\over\partial\eta}[{\cal U}_1^{ai}(x,\eta)-{\cal U}_2^{ai}(x,\eta)]
\nonumber\\
&&\hspace{-4mm}
-~2 \alpha_s
\lambda^{ai}(x,\eta)(x|U_1^\dagger{p_ip^k\over p_\perp^2}U_1-(1\leftrightarrow 2)|^{ab}
({\cal V}^\eta_1-{\cal V}^\eta_2)^b_k)\Big\}\Big) .
\nonumber
\end{eqnarray}
(When convenient, we use the notation $(...)^\eta\equiv (...)(\eta)$ for brevity).
The integral over ${\cal V}$ gives the $\delta$-function of the form
\begin{eqnarray}
&&\hspace{-3mm}
\delta(\partial_i({\partial\over\partial\eta}
[{\cal U}_1^{ai}(x,\eta)-{\cal U}_2^{ai}(x,\eta)]
\nonumber\\
&&\hspace{-3mm}
-~2i\alpha_s(x|U_1^\dagger{p_ip^k\over p_\perp^2}U_1-U_2^\dagger
{p_ip^k\over p_\perp^2}U_2|^{ab}\lambda^{b\eta}_k)
\nonumber
\end{eqnarray}
which restricts $U$ in a following way
\begin{eqnarray}
&&\hspace{-0mm}{\partial\over\partial\eta}
[{\cal U}_1^{ai}(x,\eta)-{\cal U}_2^{ai}(x,\eta)]
~\nonumber\\
&&\hspace{3mm}
=~2i\alpha_s(x|U_1^\dagger{p_ip^k\over p_\perp^2}U_1-U_2^\dagger
{p_ip^k\over p_\perp^2}U_2|^{ab}\lambda_k^{b\eta})
\label{eqnu}
\end{eqnarray}
It is convenient to rewrite Eq. (\ref{eqnu}) in the integral form (cf. Eq. \ref{tildeus}):
\begin{equation}
\hspace{-0mm}
U_i(x,\eta)=Te^{2i\alpha_st^a\!\int_{\eta_2}^\eta\! 
d\eta'(x|{p^k\over p_\perp^2}U_i(\eta')^{ab}
|\lambda_k^b(\eta'))}U_i(x_\perp,\eta_2)
\label{eqnui}
\end{equation}
where $T$ means ordering in rapidity (= our ``time''). 
The remaining integral over $\lambda$ is gaussian with 
the ``propagator'' 
\begin{equation}
\langle\lambda_i^a(x_\perp,\eta)\lambda_j^b(y_\perp,\eta')\rangle=
g_{ij}\delta^{ab}{1\over 2\alpha_s}\delta(x_\perp-y_\perp)\delta(\eta-\eta') .
\label{corrlambda}
\end{equation}

The evolution of the dipole can be represented as
\begin{eqnarray}
&&\hspace{-3mm}
U_{1y}^{\dagger\eta}
U_{1x}^{\eta}U_{2x}^{\dagger\eta} 
U_{2y}^{\eta}
~=\int D\lambda ~
e^{-\alpha_s\!\int_{\eta_2}^{\eta_1}\! d\eta\int\! d^2z_\perp\lambda^{a\eta}_{iz}
\lambda^{ai\eta}_z}~
\nonumber\\
&&\hspace{-3mm}
\times~U_{1y}^{\dagger\eta_2}
\bar{T}e^{-2i\alpha_st^a\!\int_{\eta_2}^\eta\! 
d\eta'(y|{p^k\over p_\perp^2}U_1^{\eta'}|^{ab}\lambda_k^{b\eta'})}
\nonumber\\
&&\hspace{15mm}
\times~
Te^{2i\alpha_s t^a\!\int_{\eta_2}^\eta\! d\eta'(x|{p^k\over p_\perp^2}U_1^{\eta'}
|^{ab}\lambda_k^{b\eta'})}
U_{1x}^{\eta_2}
\nonumber\\
&&\hspace{-3mm}
\times~
U_{2x}^{\dagger\eta_2} 
\bar{T}e^{-2i\alpha_st^a\!\int_{\eta_2}^\eta\! d\eta'(x|{p^k\over p_\perp^2}
U_2^{\eta'}|^{ab}\lambda_k^{b\eta'})}
\nonumber\\
&&\hspace{15mm}
\times~Te^{2i\alpha_st^a\!\int_{\eta_2}^\eta\! 
d\eta'(y|{p^k\over p_\perp^2}U^{\eta'}|^{ab}\lambda_k^{b\eta'})}
U_{2y}^{\eta_2}
\label{dipolevol}
\end{eqnarray}
where ${\bar T}$ denotes the inverse rapidity ordering.
Taking the derivative with respect to $\eta$ we get 
\begin{eqnarray}
&&\hspace{-5mm}
{\partial\over\partial\eta}U_{1y}^{\dagger\eta}
U_{1x}^{\eta}U_{2x}^{\dagger\eta} 
U_{2y}^{\eta}~=~\!\int D\lambda ~
e^{-\alpha_s\!\int_{\eta_2}^{\eta_1}\! d\eta\int\! d^2z_\perp\lambda^{a\eta}_{iz}
\lambda^{ai\eta}_z}
\nonumber\\
&&\hspace{-5mm}
\times~2i\alpha_s\Big(U_{1y}^{\dagger\eta}[t^a(x|{p^k\over p_\perp^2}
U_1^{\eta}|^{ab}\lambda_k^{b\eta})-x\leftrightarrow y]
U_{1x}^{\eta}U_{2x}^{\dagger\eta} U_{2y}^{\eta}
\nonumber\\
&&\hspace{-5mm}
-~U_{1y}^{\dagger\eta}U_{1x}^{\eta}U_{2x}^{\dagger\eta}
[t^a(x|{p^k\over p_\perp^2}
U_2^{\eta}|^{ab}\lambda_k^{b\eta})-x\leftrightarrow y] U_{2y}^{\eta}\Big)
\nonumber
\end{eqnarray}
Using the contraction
\begin{eqnarray}
&&\hspace{-5mm}
\langle\lambda_k^a(z,\eta)U(x,\eta)\rangle=-
{1\over 2}i(z|U^\dagger(\eta){p_k\over p_\perp^2}|x)^{ab}t^bU(x,\eta)
\label{withalf}\\
&&\hspace{-5mm}
\langle\lambda_k^a(z,\eta)U^\dagger(x,\eta)\rangle=-
{1\over 2}i(z|U^\dagger(\eta){p_k\over p_\perp^2}|x)^{ab}U^\dagger(x,\eta)t^b~,
\nonumber
\end{eqnarray}
one gets the non-linear evolution equation (\ref{bk}) after some algebra. 

The factor $1/2$ in the Eq. (\ref{withalf}) comes from $\theta(0)=1/2$. To avoid
this uncertainty, one should first calculate the correlations in $\lambda$ and then  differentiate with respect to rapidity (cf. Ref. \cite{pl})
\begin{eqnarray}
&&\hspace{-5mm}
{\partial\over\partial\eta}\!\int_{\eta_2}^\eta\! d\eta' d\eta"
\langle\lambda_i^a(x_\perp,\eta')\lambda_j^b(y_\perp,\eta")\rangle f(\eta')g(\eta")
\nonumber\\
&&\hspace{35mm}=-
{1\over 2\alpha_s}\delta(x-y)_\perp g_{ij}f(\eta)g(\eta)
\nonumber\\
&&\hspace{-5mm}
{\partial\over\partial\eta}\!\int_{\eta_2}^\eta\! d\eta'\int_{\eta_2}^{\eta'}\! d\eta"
\langle\lambda_i^a(x_\perp,\eta')\lambda_j^b(y_\perp,\eta")\rangle f(\eta')f(\eta")
\nonumber\\
&&\hspace{35mm}=-
{1\over 4\alpha_s}\delta(x-y)_\perp g_{ij}f(\eta)f(\eta)
\nonumber
\end{eqnarray}
The first line in the above equation should be used to make contractions between different $T$ and $\bar{T}$ in Eq. (\ref{dipolevol}) while the second line takes
care of the contractions within same $T$ or $\bar{T}$. It is easy to check that 
the result is consistent with taking $\theta(0)=1/2$ in the Eq. (\ref{withalf}).

Similarly one can demonstrate that all the hierarchy of the evolution equations for Wilson lines \cite{npb96,difope}
($\equiv$ JIMWLK equation \cite{jimwalk}) is reproduced.

\subsection{Classical equations for the Wilson-line functional integral} 
As we discussed above, the characteristic fields in the functional integral
are large but the coupling constant $\alpha_s(Q_s)$ is small due to the saturation.
In this case, we
can try to calculate the functional integral (\ref{mastegrac}) semiclassically. Using
the approximate formula 
\begin{equation}
\delta {\cal W}^a_i(U,V)\simeq 
-(W^\dagger {p_i p^j\over p_\perp^2}W)^{ab}(\delta {\cal U}_j^b+\delta {\cal V}_j^b)
\end{equation}
we get
the classical equations for the functional integral (\ref{mastegrac}) in the form
\begin{eqnarray}
&&\hspace{-2mm}
(\partial^i-ig{\cal V}_1^i)^{ab}(\dot{{\cal U}}_{1i}-\dot{{\cal U}}_{2i})^b
\label{klass}\\ 
&&\hspace{-5mm}=
2i\alpha_s(\partial^i-ig{\cal V}_1^i)^{ab}(W_F^\dagger{p^i p^j\over p_\perp^2}W_F
-W_L^\dagger{p^i p^j\over p_\perp^2}W_L)^{bc}E_j^c,
\nonumber\\ 
&&\hspace{-5mm}
(\partial^i-ig{\cal V}_2^i)^{ab}(\dot{{\cal U}}_{1i}-\dot{{\cal U}}_{2i})^b
\nonumber\\ 
&&\hspace{-5mm}=
2i\alpha_s(\partial^i-ig{\cal V}_2^i)^{ab}(W_R^\dagger{p^i p^j\over p_\perp^2}W_R
-W_B^\dagger{p^i p^j\over p_\perp^2}W_B)^{bc}E_j^c,
\nonumber\\
&&\hspace{-5mm}
(\partial^i-ig{\cal U}_1^i)^{ab}(\dot{{\cal V}}_{1i}-\dot{{\cal V}}_{2i})^b
\nonumber\\ 
&&\hspace{-5mm}=
-2i\alpha_s(\partial^i-ig{\cal U}_1^i)^{ab}(W_F^\dagger{p^i p^j\over p_\perp^2}W_F
-W_R^\dagger{p^i p^j\over p_\perp^2}W_R)^{bc}E_j^c,
\nonumber\\ 
&&\hspace{-5mm}
(\partial^i-ig{\cal U}_2^i)^{ab}(\dot{{\cal V}}_{1i}-\dot{{\cal V}}_{2i})^b
\nonumber\\ 
&&\hspace{-5mm}=
-2i\alpha_s(\partial^i-ig{\cal V}_2^i)^{ab}(W_L^\dagger{p^i p^j\over p_\perp^2}W_L
-W_B^\dagger{p^i p^j\over p_\perp^2}W_B)^{bc}E_j^c
\nonumber
\end{eqnarray}
with the initial conditions
\begin{equation}
U(\eta)=U ~{\rm at}~\eta=\eta_2,\qquad V(\eta)=V ~{\rm at}~\eta=\eta_1 .
\label{nov18}
\end{equation}
At small ${\cal V}_i$ these equations reduce to (cf. Eq. (\ref{eqnu}))
\begin{eqnarray}
&&\hspace{-2mm}
(\dot{{\cal U}}_{1i}-\dot{{\cal U}}_{2i})^a
=
2i\alpha_s(U_1^\dagger{p^i p^j\over p_\perp^2}U_1
-U_2^\dagger{p^i p^j\over p_\perp^2}U_2)^{ab}E_j^b
\nonumber\\ 
&&\hspace{-2mm}
\dot{{\cal V}}_{1i}-\dot{{\cal V}}_{2i}= O([U,V]^2)
\label{klu}
\end{eqnarray}
while in the opposite case of small ${\cal U}_i$ they are 
\begin{eqnarray}
&&\hspace{-2mm}
(\dot{{\cal V}}_{1i}-\dot{{\cal V}}_{2i})^a
=-2i\alpha_s(V_1^\dagger{p^i p^j\over p_\perp^2}V_1
-V_2^\dagger{p^i p^j\over p_\perp^2}V_2)^{ab}E_j^b
\nonumber\\ 
&&\hspace{-2mm}
\dot{{\cal U}}_{1i}-\dot{{\cal U}}_{2i}= O([U,V]^2)
\label{klv}
\end{eqnarray}
It is instructive to rewrite the equations (\ref{klu}) and (\ref{klv}) in terms of $\dot{W}$'s.
\begin{eqnarray}
&&\hspace{-5mm}
\dot{{\cal W}}_{Fi}^a-\dot{{\cal W}}_{Bi}^a=
2i\alpha_s(W_R^\dagger{p^i p^j\over p_\perp^2}W_R
-W_L^\dagger{p^i p^j\over p_\perp^2}W_L)^{ab}E_j^c
\nonumber\\ 
&&\hspace{-5mm}
\dot{{\cal W}}_{Ri}^a-\dot{{\cal W}}_{Li}^a=
2i\alpha_s(W_F^\dagger{p^i p^j\over p_\perp^2}W_F
-W_B^\dagger{p^i p^j\over p_\perp^2}W_B)^{ab}E_j^c
\nonumber\\
\label{klw}
\end{eqnarray}
From the viewpoint of the functional integral (\ref{mastegrac})
the $W$'s are the (non-local) functions 
of $U$ and $V$ variables given in the first order by Eq. (\ref{E1st}).
It would be very interesting to rewrite the Eq. (\ref{mastegral})
 of the $W$ variables themselves, that is, to construct
 the functional integral over the $W$ variables with a saddle-point
 equations given by the Eq. (\ref{klw}).

\section{Conclusion}

\begin{figure}
\includegraphics[width=0.4\textwidth]{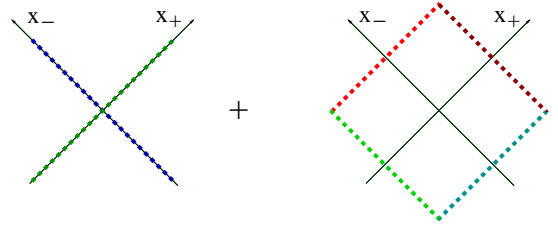}
\caption{Wilson-line structure of the effective action \label{fig:diamond1a}.}
\end{figure}

As mentioned in the Introduction, 
the popular idea of how to solve  QCD at high energies is 
to reformulate it in terms of the relevant high-energy degrees of freedom - Wilson lines.
The functional integral (\ref{mastegrac}) gives an example of such $2+1$ theory 
where $2$ stands for the transverse coordinates and $1$ for d the rapidity serving as a time variable. 
The structure of the effective action is presented in Fig. \ref{fig:diamond1a}. 
Note that the two terms in the exponent in the effective action, shown in 
Fig. \ref{fig:diamond1a}, 
are both local in $x_\perp$ but differ with respect to  the longitudinal
coordinates: the first (kinetic) term is made from the Wilson lines located at $x_+=0$ or $x_-=0$ while
the second term is made from the Wilson lines at $x_\pm=\pm\infty$. Unfortunately, the transition
between these Wilson lines is nonlocal in $x_\perp$ (see Eq. (\ref{E1st})) and so the resulting effective action 
is a non-local function of the dynamical variables $U$ and $V$.

\begin{figure}
\includegraphics[width=0.48\textwidth]{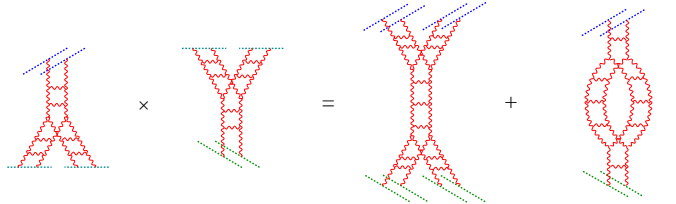}
\caption{Typical Feynman diagrams included in the effective action. \label{fig:ploops}}
\end{figure}

In should be emphasized that Eq. (\ref{mastegrac}) is only a model - the genuine effective action 
for the $2+1$ high-energy theory of Wilson lines must include all the contributions
$\sim [U,V]^n$ (as we mentioned above, an example of a $[U,V]^3$ term 
which is missing in Eq. (\ref{mastegral}) is presented in Fig. \ref{fig:3regge}). 
However, this model is correct in the case of weak projectile fields and strong target fields, and vice versa.
In terms of Feynman diagrams, the effective action (\ref{mastegrac}) includes both ``up'' and ``down''
fan ladders and the pomeron loops, see Fig. \ref{fig:ploops}. 
In the dipole language, it describes both multiplication and recombination of dipoles
(see the discussion in \cite{murecent, larecent}).
In conclusion I would like to emphasize that  the effective action  (\ref{mastegrac}) summarizes all present knowledge about the high-energy evolution of Wilson lines in a way symmetric with respect to projectile and target and hence it may serve as a  starting point for future analysis of high-energy scattering in QCD.

\begin{acknowledgments}
The author thanks  E. Iancu, L. McLerran and M. Lublinsky
 for valuable discussions and the
theory group at CEA Saclay for kind hospitality. 
This work was supported by contract
DE-AC05-84ER40150 under which the Southeastern Universities Research
Association (SURA) operates the Thomas Jefferson National Accelerator
Facility.
\end{acknowledgments}

\appendix

\section{Classical fields in the first order in $[U,V]$\label{sect:a1}}
Following Ref. \cite{prd04}, we take the zero-order approximation in the form of the sum of 
the two shock waves (\ref{shoksum})
\begin{eqnarray}
&&\hspace{-5mm}\bar{A}_i^{(0)}=U_i\theta(-x_\ast)+V_i\theta(-x_\bullet),~~~
\bar{A}_\bullet^{(0)}=\bar{A}_\ast^{(0)}=0
 \label{shoksum1}
\end{eqnarray}
We will expand the ``deviation'' of the full QCD solution from
the QED-type ansatz (\ref{shoksum1}) in powers of commutators $[U,V]$. 
To carry this out, we
 shift $A\rightarrow A+\bar{A}_i^{(0)}$ in the functional integral 
(\ref{mastegral})
and obtain
\begin{eqnarray}
&&\hspace{0mm} 
\int\! DA~
 \exp\Big\{i\!\int\! d^4z(
 {1\over 2} A^\mu \bar{D}_{\mu\nu}A^\nu+gT^\mu A_\mu)\Big\}.
 \label{mastegral2}
\end{eqnarray}
Here $D_{\mu\nu}=D^2(\bar{A})g_{\mu\nu}-2i \bar{F}_{\mu\nu}$
is the inverse propagator in the background-Feynman gauge
and
$T_\mu$ is the linear term for the trial configuration
(\ref{shoksum1}). Since the only non-zero component 
of the field strength for the ansatz (\ref{shoksum1}) is
\begin{eqnarray}
\bar{F}^{(0)}_{ik}&=&-i[U_{1i},V_{1k}]\theta(F)
-i[U_{1i},V_{2k}]\theta(R)
\label{efnol}\\
&-&
i[U_{2i},V_{1k}]\theta(L)
-i[U_{2i},V_{2k}]\theta(B)-(i\leftrightarrow k)
\nonumber
\end{eqnarray}
the linear term $T_{\mu}=\bar{D}^\rho \bar{F}^{(0)}_{\rho\mu}$ is
\begin{eqnarray}
&&\hspace{-5mm}T_\ast~=~T_\bullet~=~0,
\label{Tsap}\\
&&\hspace{-5mm}
T^i~=~(i\partial_k+g[U_{1k}+V_{1k},)([U_1^i,V_1^k]-i\leftrightarrow k)\theta(F)
\nonumber\\
&&\hspace{-5mm}
+~(i\partial_k+g[U_{1k}+V_{2k},)([U_1^i,V_2^k]-i\leftrightarrow k)\theta(R)
\nonumber\\
&&\hspace{-5mm}
+~(i\partial_k+g[U_{2k}+V_{2k},)([U_2^i,V_1^k]-i\leftrightarrow k)\theta(L)
\nonumber\\
&&\hspace{-5mm}
+~(i\partial_k+g[U_{2k}+V_{2k},)([U_2^i,V_2^k]-i\leftrightarrow k)\theta(B)
\nonumber
\end{eqnarray}
where $\theta(F)\equiv\theta(z_\ast)\theta(z_\bullet)$, $\theta(R)\equiv\theta(z_\ast)\theta(-z_\bullet)$, 
$\theta(L)\equiv\theta(-z_\ast)\theta(z_\bullet)$, and 
 $\theta(B)\equiv\theta(-z_\ast)\theta(-z_\bullet)$

Expanding in powers of $T$ in the functional integral (\ref{mastegral2}) 
one gets the set of Feynman diagrams in the external fields (\ref{shoksum}) with the sources
(\ref{Tsap}). The parameter of the expansion is $g^2[U_i,V_j]$ ($\sim [U,V]$, see Eq.
(\ref{defui})). 

The general formula for the classical solution in the first order in $[U,V]$ has the form
\begin{eqnarray}
\hspace{0mm}
\bar{A}^{(1)a}_\mu(x) &=&
ig\!\int\! d^4z
\langle A^a_\mu(x)A^{b\nu}(z)\rangle_{\bar{A}} T^b_\nu(z)
\label{1order}
\end{eqnarray}
The Green functions in the background of the Eq. (\ref{shoksum})
field can be approximated by cluster expansion
\begin{eqnarray}
&&\hspace{0mm}
\langle A_\mu(x)A^\nu(z)\rangle_{\bar{A}}
\nonumber \\
&&\hspace{0mm} =~ 
\langle A_\mu(x)A^\nu(z)\rangle_U
+\langle A_\mu(x)A^\nu(z)\rangle_V
\nonumber \\
&&\hspace{0mm}-~
\langle A_\mu(x)A^\nu(z)\rangle_0+O([U,V])
\label{cluster}
\end{eqnarray}
where $\langle A_\mu(x)A^\nu(z)\rangle_0$
is the perturbative propagator  and
\begin{eqnarray}
&&\hspace{-3mm}
\langle A^a_\mu(x)A^b_\nu(y)\rangle_U
\stackrel{x_\ast>0, y_\ast<0}{~=~}\!\int\! dz \delta({2\over s}z_\ast)~\Big\{U^\dagger_{1x}
(x|{1\over p^2+i\epsilon }|z)
\nonumber\\
&&\hspace{-3mm}
\times~
\Big(2\alpha
g_{\mu\nu}U_1U_2^\dagger 
+{4i\over s}
(p_{2\nu}\partial_\mu(U_1U_2^\dagger)_z+\mu\leftrightarrow\nu)
\nonumber\\
&&\hspace{-3mm}
-~{4p_{2\mu}p_{2\nu}\over \alpha s^2}\partial_\perp^2(U_1U_2^\dagger )_z
\Big)(z|{1\over p^2+i\epsilon }|y)U_{2y}\Big)^{ab}
\label{greenfun}
\end{eqnarray}
is the propagator in the background of the shock wave $U$ (the propagator 
in the $V$ background is obtained by the replacement $U\leftrightarrow V$, 
$p_2\leftrightarrow p_1$.

Substituting Eq. (\ref{greenfun}) and  (\ref{Tsap}) into the above equation, one 
obtains (the details of the calculations can be found in Ref. \cite{prd04} and here we present only the the 
final set of gauge fields): 
\begin{eqnarray}
A^\mu&=&\theta(F)\Big\{{\cal W}^{\mu \perp}_F(x_\perp)
-gt^a (W_F^\dagger{1\over\partial^2-i\epsilon}W_F)^{ab} L_F^{\mu b}(x)
\Big\}
\nonumber\\
&+& (F\leftrightarrow L)+ (F\leftrightarrow R)+ (F\leftrightarrow B)
\label{fiilds}
\end{eqnarray}
where $L_F^i~=~L_L^i~=~L_R^i~=~L_B^i~=~2E^i$ and
\begin{eqnarray}
\hspace{-5mm}
L_{F\ast}&=&
2[V_{1i}-V_{2i},{1\over\beta+i\epsilon}E_R^i
-{1\over\beta-i\epsilon}E_B^i]
\nonumber\\
L_{F\bullet}&=&2[U_{1i}-U_{2i},{1\over\alpha+i\epsilon}E_L^i
-{1\over\alpha-i\epsilon}E_B^i]
\nonumber\\
L_{L\ast} &=&    
2[V_{1i}-V_{2i},{1\over\beta+i\epsilon}E_R^i
-{1\over\beta-i\epsilon}E_B^i]
\nonumber\\
L_{L\bullet}&=&2[U_{1i}-U_{2i},{1\over\alpha+i\epsilon}E_F^i
-{1\over\alpha-i\epsilon}E_R^i]
\nonumber\\
L_{R\ast}&=&2[V_{1i}-V_{2i},
{1\over\beta+i\epsilon}E_F^i-{1\over\beta-i\epsilon}E_L^i]
\nonumber\\
L_{R\bullet} &=&   2[U_{1i}-U_{2i},{1\over\alpha+i\epsilon}E_L^i
-{1\over\alpha-i\epsilon}E_B^i]
\nonumber\\
L_{B\ast}&=&2[V_{1i}-V_{2i},
{1\over\beta+i\epsilon}E_F^i-{1\over\beta-i\epsilon}E_L^i]
\nonumber\\
L_{B\bullet} &=& [U_{1i}-U_{2i},{1\over\alpha+i\epsilon}E_F^i
-{1\over\alpha-i\epsilon}E_R^i]
\label{els}
\end{eqnarray}
where ${2/s\over\alpha\pm i\epsilon}{\cal O}(x)\equiv i \int_0^{\pm\infty} du {\cal O}(x+up_2)$
and ${2/s\over\beta\pm i\epsilon}{\cal O}(x)\equiv i \int_0^{\pm\infty} du {\cal O}(x+up_1)$.
It is easy to check the background-Feynman gauge condition $(i\partial_\mu +g[{\cal W}_F^\mu,)L_{F\mu}=0$ (and similarly for three other quadrants of the space). 

The transverse part $E_i$ agrees with the results  of Sec. \label{sect:effactc} while
the longitudinal part (\ref{els}) does not literally agree with (\ref{Ts}) (see the footnote
after that equation). It should be emphasized that, unlike the calculations with trial configuration ({\ref{fild}), the Feynman diagrams in the background of the 
ansatz (\ref{shoksum1}) are free from uncertainties like $\theta(0)$.

Let us rederive now the effective action (\ref{withslope}) starting from the ansatz 
(\ref{shoksum1}) and the fields (\ref{fiilds}). 
Since the only non-zero component 
of the field strength for the ansatz (\ref{shoksum1}) is transverse (see Eq. (\ref{efnol}),
we have
\begin{eqnarray}
&&\hspace{-5mm}
S_{\rm eff}~=~-{1\over 4}\!\int\! d^4z 
\bar{F}^{(0)a}_{ik}\bar{F}^{(0)a,ik}
\nonumber\\
&&\hspace{-5mm}
+~{i\over 2}\!\int\! d^4z d^4z' T^a_i(z)T^b_j(z')
\langle A^{ai}(z)A^{bj}(z')\rangle
\nonumber\\
&&\hspace{-5mm}
=~-{1\over 4}\!\int\! d^4z \Big(
\bar{F}^{(0)a}_{ik}\bar{F}^{(0)a,ik}
\nonumber\\
&&\hspace{-5mm}
+~i\!\int\! d^4z' 
\bar{F}^{(0)a}_{ik}(z)
\langle (\bar{D}^iA^{ak}(z)- i\leftrightarrow k)A^{bj}(z')\rangle T^b_j(z')\Big\}
\nonumber\\
&&\hspace{-5mm}
=~-{1\over 4}\!\int\! d^4z 
\bar{F}^{(0)a}_{ik}\bar{F}^{(1)a,ik}
\label{s1}
\end{eqnarray}
where $\bar{F}^{(1)a,ik}$ is a field strength in the first order in $[U,V]$. Using 
the fields (\ref{fiilds}) we obtain
\begin{eqnarray}
\hspace{-2mm}
F^{(1)a}_{ik}(z) &=&-2g\theta(F)(z|W_F^\dagger{\partial_i\over \partial^2-i\epsilon}W_F|^{ab}0,E_k^b)
\label{efiks}\\
 &+&
 (F\leftrightarrow L)+ (F\leftrightarrow R)+ (F\leftrightarrow B) -(i\leftrightarrow k)
\nonumber
\end{eqnarray}
(where $|0, E_i)\equiv \int d^2z'_\perp |0,z'_\perp)E_i(z'_\perp)$)
and therefore
\begin{eqnarray}
&&\hspace{-5mm}
S_{\rm eff}~
\label{s3}\\
&&\hspace{-5mm}
=~-ig^2\!\!\int\! d^4z 
\Big\{\theta(F)([U_1^i,V_1^k]- i\leftrightarrow k)^a
 (W_F^\dagger{\partial_i\over\partial^2}W_F)^{ab} E_k^b
\nonumber\\
&&\hspace{10mm}
+~\theta(R)([U_1^i,V_2^k]- i\leftrightarrow k)^a
 (W_R^\dagger{\partial_i\over\partial^2}W_R)^{ab} E_k^b
\nonumber\\
&&\hspace{10mm}
+~\theta(L)([U_2^i,V_1^k]- i\leftrightarrow k)^a
 (W_L^\dagger{\partial_i\over\partial^2}W_L)^{ab} E_k^b
\nonumber\\
&&\hspace{10mm}
+~\theta(L)([U_2^i,V_2^k]- i\leftrightarrow k)^a
 (W_B^\dagger{\partial_i\over\partial^2}W_B)^{ab} E_k^b\Big\}
\nonumber
\end{eqnarray}
A typical integral in the above equation has the form
\begin{eqnarray}
&&\hspace{-10mm}
\int\! d^4z d^2z'_\perp~\theta(z_\ast)\theta(z_\bullet)
 f(z_\perp)(z|{p_i\over p^2+i\epsilon}|0,z'_\perp)g(z'_\perp)
 \nonumber\\
&&\hspace{-10mm}
={i\over 2\pi}\!\int_0^\infty\! {d\alpha\over\alpha}\!\int\! d^2z_\perp d^2z'_\perp
f(z_\perp)(z_\perp|{p_i\over p_\perp^2}|z'_\perp)g(z'_\perp)
\label{s6}
\end{eqnarray}
In the LLA, the integral $\!\int_0^\infty\! {d\alpha\over\alpha}$ is replaced by 
${1\over 2}\Delta\eta$. More accurately, one should remember that the slopes of 
Wilson lines are $e_1=p_1+e^{-\eta_1}p_2$ and 
$e_2=p_2+e^{\eta_2}p_1$ as shown in Eq. (\ref{fak2times}). In this case, 
$\theta(z_\ast)\theta(z_\bullet)$ in the integrand of Eq. (\ref{s6}) will be replaced
by $\theta(z_\ast+e^{\eta_2}z_\bullet)\theta(z_\bullet+e^{-\eta_1}z_\ast)$ so one obtains
\begin{eqnarray}
&&\hspace{-5mm}
\int\! d^4z d^2z'_\perp~\theta(z_\ast)\theta(z_\bullet)
 f(z_\perp)(z|{p_i\over p^2+i\epsilon}|0,z'_\perp)g(z'_\perp)
\nonumber\\
&&\hspace{-5mm}
=~-\!\int\! d^2z_\perp d^2z'_\perp f(z_\perp)g(z'_\perp)\!\int\! {d\alpha d\beta
d^2p_\perp\over 16\pi^4}~e^{i(p,z-z')_\perp}
 \nonumber\\
&&\hspace{-5mm}\times~
{p_i\over \alpha\beta s-p_\perp^2+i\epsilon}~
{1\over(\alpha+e^{\eta_2}\beta-i\epsilon)(\beta+e^{-\eta_1}\alpha-i\epsilon)}
 \nonumber\\
&&\hspace{-5mm}
=~-i\!\int\! d^2z_\perp d^2z'_\perp f(z_\perp)g(z'_\perp)\!\int_0^\infty\! {d\alpha \over\alpha}
{d^2p_\perp\over 8\pi^3}~e^{i(p,z-z')_\perp}
 \nonumber\\
&&\hspace{-5mm}\times~
\Big({p_i\over e^{-\eta_1} \alpha^2 s+p_\perp^2}-
{p_i\over e^{-\eta_2} \alpha^2 s+p_\perp^2}\Big)
 \nonumber\\
&&\hspace{-5mm}
=~-{i\over 4\pi}\Delta\eta
\!\int\! d^2z_\perp d^2z'_\perp f(z_\perp)(z_\perp|{p_i\over p_\perp^2}|z'_\perp)g(z'_\perp) 
\label{s7}
\end{eqnarray}
where $\Delta\eta=\eta_1-\eta_2$. Performing the integrations over $z_\ast, z_\bullet$ in Eq. 
(\ref{s3}) we get
\begin{eqnarray}
&&\hspace{-5mm}
S_{\rm eff}
\label{s8}\\
&&\hspace{-5mm}
=~-\alpha_s\Delta\eta\!\int\! d^2z_\perp  
\Big\{([U_1^i,V_1^k]- i\leftrightarrow k)^a
 (W_F^\dagger{\partial_i\over\partial_\perp^2}W_F)^{ab}
\nonumber\\
&&\hspace{10mm}
-~([U_1^i,V_2^k]- i\leftrightarrow k)^a
 (W_R^\dagger{\partial_i\over\partial_\perp^2}W_R)^{ab} 
\nonumber\\
&&\hspace{10mm}
-~([U_2^i,V_1^k]- i\leftrightarrow k)^a
 (W_L^\dagger{\partial_i\over\partial_\perp^2}W_L)^{ab}
\nonumber\\
&&\hspace{10mm}
+~([U_2^i,V_2^k]- i\leftrightarrow k)^a
 (W_B^\dagger{\partial_i\over\partial_\perp^2}W_B)^{ab}\Big\} E_k^b
\nonumber\\
&&\hspace{-5mm}
=~-i\alpha_s\Delta\eta\!\int\! d^2z_\perp~  E_i^a E^{ai}
\nonumber
\end{eqnarray}
which coincides with Eq. (\ref{withslope}).

Finally, let us demonstrate that the ``diamond'' trace of four (non-differentiated) Wilson lines 
is trivial (this is related to the fact that the field strength $F_{+-}$ vanishes in the leading order, see Eqs. (\ref{fiilds}) and (\ref{els})). To regularize the corresponding expressions, we consider the
``original'' tilted Wilson loop shown in Fig. \ref{fig:diamond2} for the finite size $L$.
\begin{figure}
\includegraphics[width=0.25\textwidth]{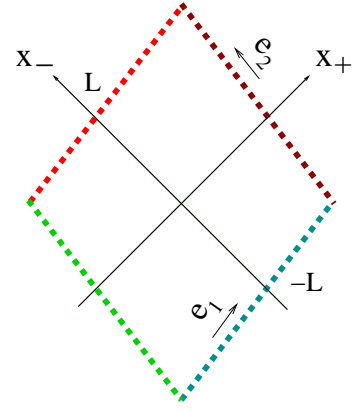}
\caption{Trace of four Wilson lines. \label{fig:diamond2}}
\end{figure}
We need to prove that
\begin{eqnarray}
&&\hspace{-8mm}
\lim_{L\rightarrow\infty} {\rm tr}\{[-L e_1+Le_2+x_\perp,L e_1+Le_2+x_\perp]
\label{barediamond}\\
&&\hspace{3mm}\times~
[L e_2+L e_1+x_\perp,-L e_2+L e_1+x_\perp]
\nonumber\\
&&\hspace{3mm}\times~
[L e_1-L e_2+x_\perp,-L e_1-L e_2+x_\perp]
\nonumber\\
&&\hspace{3mm}\times~[-L e_2-L e_2+x_\perp,L e_2-L e_2+x_\perp]\}~=~1
\nonumber
\end{eqnarray}
in the leading nontrivial order in $[U,V]$ .

Consider the case ${\cal U}_i\ll 1$, ${\cal V}_i\sim1$ (the opposite case ${\cal V}_i\ll 1$, ${\cal U}_i\sim1$
is similar).  It is easy to see from Eq. (\ref{els}) that $[L e_2\pm Le_1+x_\perp,-L e_2\pm Le_1+x_\perp]\sim [U,V]^2$ so we are left
with 
\begin{eqnarray}
&&\hspace{-3mm}
\lim_{L\rightarrow\infty} {\rm tr}\{[-L e_1+Le_2+x_\perp,L e_1+Le_2+x_\perp]
\nonumber\\
&&\hspace{-3mm}
\times~[L e_1-L e_2+x_\perp,-L e_1-L e_2+x_\perp]\}
\label{twolines}
\end{eqnarray}
At this point, we can take the limit $L\rightarrow\infty$ in the $e_1$ direction. We obtain:
\begin{eqnarray}
&&\hspace{-3mm}
[x_\perp+L e_2,x_\perp+L e_2+\infty e_1]
\label{nextolasteqn}\\
&&\hspace{13mm}
\times~[x_\perp-L e_2+\infty e_1,x_\perp-L e_2]-1
\nonumber\\
&&\hspace{-3mm}
=~{i\over \pi^2}\!\int\! d\alpha d\beta~
{\sin\alpha L\over(\alpha+e^{\eta_2}\beta-i\epsilon)(\beta+e^{-\eta_1}\alpha-i\epsilon)}
\nonumber\\
&&\hspace{-3mm}
\times~
(x_\perp|U_1^\dagger{1\over\alpha\beta s-p_\perp^2+i\epsilon}U_1|^{ab}|[U_{1i}-U_{2i},E_L^i-E_2^i]^b)
\nonumber\\
&&\hspace{-3mm}=
-{2\over\pi}\!\int_0^\infty\! d\alpha~{\sin\alpha L\over\alpha}
(x_\perp|U_1^\dagger
\Big({1\over e^{-\eta_1} \alpha^2 s+p_\perp^2}
\nonumber\\
&&\hspace{3mm}
-~{1\over e^{-\eta_2} \alpha^2 s+p_\perp^2}\Big)
U_1|^{ab}|[U_{1i}-U_{2i},E_L^i-E_2^i]^b)
\nonumber
\end{eqnarray}
We see now that in the limit $L\rightarrow\infty$ the r.h.s. of Eq. (\ref{lasteqn}) 
vanishes so the l.h.s. is at best $\sim [U,V]^2$ . Similarly,
\begin{eqnarray}
&&\hspace{-3mm}
\lim_{L\rightarrow\infty}[x_\perp+L e_2,x_\perp+L e_2-\infty e_1]
\label{lasteqn}\\
&&\hspace{13mm}
\times~[x_\perp-L e_2-\infty e_1,x_\perp-L e_2]~=~1
\nonumber
\end{eqnarray}
 and therefore the trace (\ref{twolines}), which is product of  l.h.s. of Eq. (\ref{nextolasteqn})
 and Eq.  (\ref{lasteqn}), is equal to 1 in the leading order.

\widetext
\mbox{}

\end{document}